\documentclass[5p]{elsarticle}
\usepackage[switch,modulo,mathlines]{lineno}
\usepackage{hyperref}
\usepackage{amssymb}
\usepackage{amsmath}
\journal{Astronomy and Computing}
\bibpunct{(}{)}{;}{a}{}{,}  
\usepackage{nicefrac}	    
\usepackage{microtype}      
\usepackage{listings}
\usepackage{color}
\definecolor{mygreen}{rgb}{0.3,0.5,0}
\definecolor{myred}{rgb}{0.8,0,0}
\definecolor{mygray}{rgb}{0.5,0.5,0.5}

\lstdefinestyle{cpp}{
  language=C++,
  frame=tb,
  aboveskip=1mm,
  belowskip=1mm,
  columns=flexible,
  numbers=left,
  firstnumber=1,
  numberstyle={\tiny\color{mygray}},
  basicstyle={\footnotesize\ttfamily},
  stringstyle=\color{myred},
  keywordstyle=\color{mygreen},
  showstringspaces=false,
  keepspaces=true, 
  tabsize=4
}

\newcommand{\cpp}[1]{{\small\ttfamily #1}}

\begin{document}
\begin{frontmatter}
\title{SKIRT~9: redesigning an advanced dust radiative transfer code to allow\\kinematics, line transfer and polarization by aligned dust grains}

\author{Peter Camps\corref{correspondingauthor}}
\ead{peter.camps@ugent.be}
\cortext[correspondingauthor]{Corresponding author}
\author{Maarten Baes\corref{none}}
\address{Sterrenkundig Observatorium, Universiteit Gent, Krijgslaan 281, B-9000 Gent, Belgium}

\begin{abstract}
The open source SKIRT Monte Carlo radiative transfer code has been used for more than 15 years to model the interaction between radiation and dust in various astrophysical systems. In this work, we present version 9 of the code, which has been substantially redesigned to support long-term objectives. We invite interested readers to participate in the development, testing and application of new features such as including gas media types in addition to dust, performing line transfer in addition to continuum radiation transfer, and modeling polarization by non-spherical dust grains aligned by magnetic fields. We describe the major challenges involved in preparing the code for these and other extensions, as well as their resolution, including a completely new treatment of wavelengths to support kinematics. SKIRT~9 properly runs over 400 handcrafted functional tests and successfully performs all relevant benchmarks. The source code and all documentation is publicly available for use and ready for further collaborative development.
\end{abstract}

\end{frontmatter}


\section{Introduction}
\label{Introduction}

Properly tracking the complex interplay between radiation and the interstellar medium in astrophysical objects is gaining importance as the quality of both observations and computer models continues to increase. For example, recent cosmological simulations of galaxy formation such as EAGLE \citep{Crain2015, Schaye2015}, MUFASA \citep{Dave2016,Dave2019} and IllustrisTNG \citep{Pillepich2019, Nelson2019}, and zoom-in simulations such as Auriga \citep{Grand2017} and FIRE-2 \citep{Hopkins2018} reproduce structure on sufficiently small spatial scales for the precise geometry to have a significant impact on the global radiation signature \citep[see, e.g.,][]{Witt1992, Baes2001, Saftly2015}. Properly comparing the results of these simulations with observations requires a detailed three-dimensional radiation transfer analysis taking into account the effects of dust on the stellar radiation. This calculation is usually performed as a post-processing step starting from a specific snapshot produced by the hydro-dynamical simulation \citep[see, e.g.,][]{Calore2015, Bignone2016, Camps2016, Santos2017, Guidi2018, Lahen2018, Behrens2018, Ma2019, Vogelsberger2019}.

The SKIRT code (ascl:1109.003) has been developed for these purposes. It uses the Monte Carlo technique \citep{Whitney2011, Steinacker2013} to trace the effects of scattering, absorption and emission by dust, and offers a library of built-in geometry models as well as the capability of importing models from hydro-dynamical simulation snaphots. The code was initially introduced by \citet{Baes2003,Baes2011} and has since been re-architected and outfitted with a user-friendly interface as presented by \citet{Camps2015a}. Various other aspects of the code are described elsewhere, e.g., advanced spatial grids \citep{Saftly2013, Saftly2014, Camps2013}, input model geometries \citep{Baes2015}, dust heating \citep{Camps2015b}, polarization \citep{Peest2017} and parallelization \citep{Verstocken2017}. Recent versions of the code have successfully been applied by various research teams around the world to galaxies \citep[see, e.g.,][]{Calore2015, Mosenkov2016, Bignone2016, Camps2016, Viaene2017, Trayford2017, Lahen2018, Behrens2018, Rodriguez2019, Ma2019, Williams2019, Vogelsberger2019}, active galactic nuclei \citep{Stalevski2016, Stalevski2017, Stalevski2019}, circumstellar disks \citep{Deschamps2015, Hendrix2016}, and molecular clouds \citep{Hendrix2015, Monceau2017}.

With its advanced handling of dust continuum radiation, SKIRT is applicable to a wide range of science cases. On the other hand, it is often desirable to study other observational signatures, such as gas emission and absorption lines, alongside the dust continuum fluxes. On the modeling side, this may be accomplished by combining the results of multiple codes, each dedicated to solving a specific aspect of the problem \citep[see, e.g.,][]{Behrens2019}. Simulating these various signatures in a single code, however, would simplify the modeling effort and would enable self-consistent calculation of all synthetic-observational results.

This is why we set out to adjust the SKIRT framework so that it can easily support new media types and new physical processes beyond dust continuum radiation transfer. We realized early on that the core SKIRT team lacks the expertise and the resources to actually implement many of these physical processes on top of the framework. We therefore intend this to be a collaborative effort, involving research teams with the relevant know-how to help develop and test new features, and to apply them to relevant science cases. In support of this goal, we have increased our efforts to provide a proper open-source collaboration environment.

In this paper, we discuss the transition from SKIRT version 8 to version 9, representing a substantial redesign of the code to allow including gas media types in addition to dust, performing line transfer in addition to continuum radiation transfer, and modeling polarization by non-spherical dust grains aligned by magnetic fields. We also describe a graphical user interface feature that was actually introduced before SKIRT~9 but has not been presented elsewhere (see Sect.~\ref{MakeUp}).

In Sect.~\ref{Objectives} we present the goals formulated at the start of the redesign process and we consider the resulting challenges.
In Sect.~\ref{Choices} we present the design choices made in the major areas of the code in response to the design goals. 
In Sect.~\ref{Features} we provide more detail on the design and operation of specific features, focusing on novel design mechanisms and heuristics.
In Sect.~\ref{Performance} we discuss the tests employed to verify correctness of the results and we present performance comparisons to earlier SKIRT versions. 
In Sect.~\ref{Summary} we summarize and provide a future outlook. 

Just as for earlier versions, the new SKIRT~9 code is hosted on GitHub\footnote{https://github.com/SKIRT/SKIRT9} and the documentation is published on the SKIRT web site,\footnote{http://www.skirt.ugent.be} including user guides and tutorials as well as reference material.


\section{Objectives}
\label{Objectives}

\subsection{Long term goals}

As discussed in the introduction, in the long run we would like to mold SKIRT into a code that handles a broader range of physics, including:
\begin{itemize}
\item multiple media types, i.e.\ dust, electrons, and gas (or at least hydrogen in its various forms) in a single simulation;
\item kinematic effects from moving sources and media;\footnote{The implementation of kinematics in early SKIRT versions \citep{Baes2002,Baes2003} was removed later on.}
\item line transfer as well as continuum transfer, including scattering, absorption, and re-emission;
\item self-consistent medium state calculation, such as for gas ionization or dust destruction;
\item polarized radiation, including the effects of spheroidal dust grains that are partially aligned along magnetic field lines.
\end{itemize}

At the same time, we wish to maintain and further develop the existing key capabilities, including:
\begin{itemize}
\item importing models from different types of hydro-dynamical simulation snapshots (using smoothed particles, hierarchical grids, or unstructured grids);
\item built-in models for spatial density distributions (geometries), source spectra, material properties, an-isotropic emission profiles, and more;
\item user-friendly run-time configuration to generically combine all of these options.
\end{itemize}

\subsection{Challenges}
\label{Challenges}

These ambitious goals have some fundamental consequences for the structure of the SKIRT code. Most importantly, the treatment of wavelengths in SKIRT~8 (and prior versions) can no longer be maintained. A SKIRT~8 configuration file defines a single wavelength grid that is used throughout the simulation. The grid specifies a fixed list of wavelengths that can be assigned to photon packets and on which optical material properties are (re-)sampled, and it defines the wavelength bins used for tracking the radiation field and for recording output fluxes. Because the same grid is being used globally (for a particular simulation), the implementation can assign an integer index to each wavelength and accessing wavelength-dependent quantities requires just a trivial indexing operation.

Returning to the objectives for SKIRT~9, gas line transfer requires support for kinematics because absorption and scattering cross sections may vary significantly over the wavelength Doppler-shifts caused by the relative motion of sources and media. This in turn requires tracking a precise, variable wavelength while a photon packet moves through the medium, as opposed to the fixed wavelength index of the SKIRT~8 approach. Furthermore, it becomes desirable to maintain the original discretization grid for tabulated quantities such as source spectra and optical properties rather than re-sampling them onto some coarser grid, because interpolating from the finer original grid will be more accurate.

A second important challenge is to handle the growing complexity of the input model. Supporting kinematics requires specifying velocities across the spatial domain. Similarly, in addition to changes in the photon cycle, supporting spheroidal dust grains requires specifying the direction and degree of grain alignment across the spatial domain. The SKIRT~8 geometry and decorator system has been designed to model a density field normalized to unit mass \citep{Baes2015}. Generalizing these classes to also model vector fields (for velocities and directions) and unnormalized scalar fields (for alignment degrees) seems essentially impossible, not in the least because the set of relevant models would differ greatly between the various use cases. A more pragmatic approach is called for, preferably one without too much loss of generality.

Finally, once the appropriate framework has been put in place and the implementation of new physics is being initiated, the potential application domain for the code will expand into new areas outside of the core expertise of our research group. This poses a third major challenge: turning the SKIRT project into a truly collaborative effort. Contributions to the code from outside the core SKIRT team have been limited, although the code and its documentation have been publicly available for several years \citep{Camps2015a}, and over 60 refereed astrophysical papers have been published that include results generated by SKIRT (see the `Publications' section on the SKIRT web site). It is clear that we need to do more to achieve our goals in this area.


\section{Design choices}
\label{Choices}

\begin{figure}
  \centering
  \includegraphics[width=0.75\columnwidth]{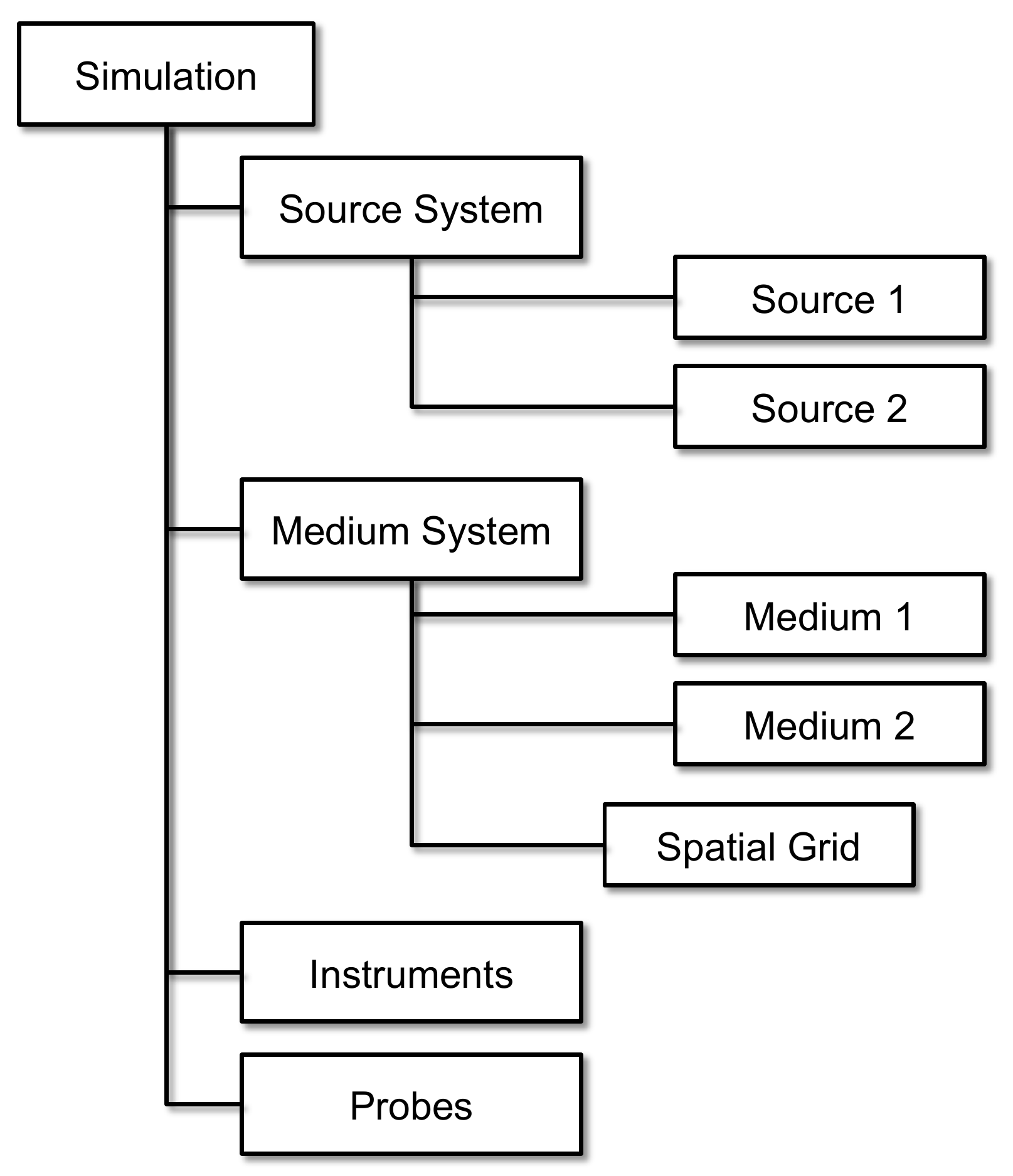}
  \caption{An overview of the top-level items to be configured for a typical SKIRT~9 simulation.}
  \label{fig:configuration}
\end{figure}

In this section we describe how we have attempted to address the challenges discussed in Sect.~\ref{Challenges} by redesigning some major areas of the SKIRT code for version 9. In the next section, Sect.~\ref{Features}, we present some of these design elements in much more depth. To help guide the discussion, Fig.~\ref{fig:configuration} offers an overview of the major items to be configured for a typical SKIRT~9 simulation.

\subsection{Wavelengths}
\label{Wavelengths}

\begin{figure*}
  \centering
  \includegraphics[width=0.99\textwidth]{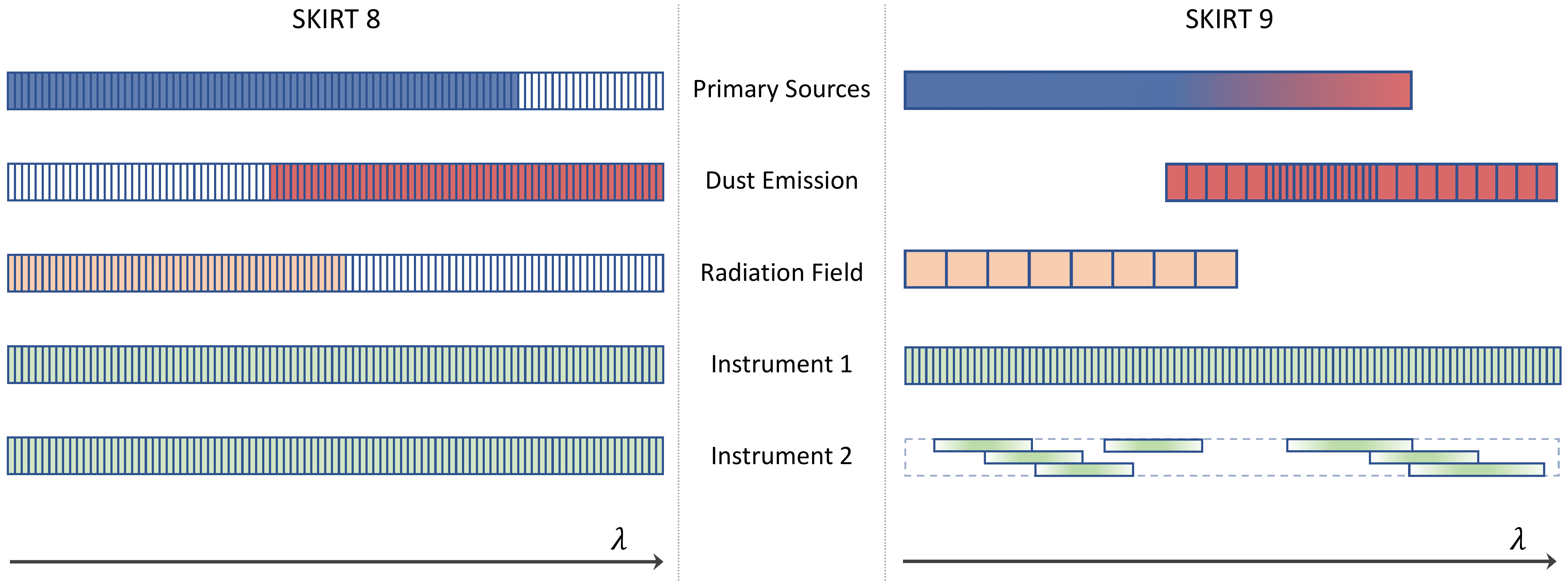}
  \caption{The treatment of wavelengths before and after the transition discussed in Sect.~\ref{Wavelengths} for various areas of the simulation (see labels in central column). In SKIRT~8 (left), all wavelength grids in a simulation are identical and a photon packet has a wavelength that is fixed at one of the wavelength bin centers. As a result, the computing resources associated with bins in unused wavelength ranges are wasted (see the white bins in the figure). In SKIRT~9 (right), wavelength grids for different purposes are fully uncoupled, and arbitrary photon packet wavelengths are sampled from the continuous spectral distribution. Instruments can use possibly overlapping broadband wavelength bins in addition to regular adjacent bins.}
  \label{fig:wavelengths89}
\end{figure*}

As indicated in Sect.~\ref{Challenges}, one of the most fundamental changes in SKIRT~9 is in the treatment of wavelengths. This change requires updating a core concept, the photon packet, and has implications in many other areas of the code. 

The key properties of a photon packet now include its wavelength and its weight. The wavelength property specifies the wavelength (or equivalently, the frequency) of all photons in the packet. The weight property specifies the number of photons carried by the packet, or more precisely the number of photons per unit of time (because SKIRT solves the time-independent radiation transfer equation). At launch, a photon packet receives a wavelength sampled from the source spectrum and a luminosity, i.e.\ its share of the total luminosity of the source. The wavelength is stored as given. The luminosity is converted to a weight (number of photons) for storage in the photon packet.\footnote{We use the term \emph{luminosity} to indicate energy per unit time \emph{carried} by a photon packet, although, strictly speaking, it should only be used to indicate energy per unit time \emph{emitted} by a source.}

During a photon packet's life cycle, updates can occur to its weight, e.g.\ because of biasing \citep[see, e.g.,][]{Baes2016}, and its wavelength, e.g.\ after being scattered by a moving medium (see next paragraph in this section). Because these updates can be fractional, both wavelength and weight are stored as floating-point values. Adjusting a photon packet's wavelength indirectly affects the luminosity represented by the packet, because the latter is directly proportional to the frequency and thus inversely proportional to the wavelength.

The wavelength of a photon packet is defined relative to the model coordinate system. In other words, a medium at rest relative to the model coordinate system sees this wavelength. Velocities of sources and media are also defined relative to the model coordinate system. Instruments are considered to be at rest relative to the model coordinate system. When a photon packet is launched, its wavelength is Doppler shifted according to the component of the source velocity in the photon packet's direction. When a photon packet interacts with a medium, the perceived wavelength is derived by Doppler shifting the packet's wavelength according to the component of the medium velocity in the photon packet's incoming direction. Specifically, registration of a photon packet's contribution to the radiation field uses this perceived wavelength. After a scattering interaction, the photon packet's wavelength is replaced by the perceived wavelength, Doppler shifted according to the component of the medium velocity in the photon packet's outgoing direction.

SKIRT~9 does not use a simulation-wide wavelength grid. Instead, there are multiple independent wavelength grids, each specialized for a particular purpose (see Fig.~\ref{fig:wavelengths89}). For example:
\begin{itemize}
\item Wavelength-dependent material properties are tabulated on some private wavelength grid, possibly depending on the resolution requirements of each material type.
\item For each instrument, the user configures the wavelength grid that will be used for binning detected photon packets. To avoid repetition, a default instrument wavelength grid can be configured as well.
\item The radiation field wavelength grid is used to register photon packet contributions to the mean radiation field in each spatial cell.
\item The dust emission grid is used to store the dust emission spectra calculated “on the fly” for each cell during secondary emission. 
\end{itemize}
Apart from offering extra flexibility, this setup often substantially reduces memory requirements, as discussed in Sects.~\ref{WaveConfig} and \ref{Memory}.

Furthermore, each radiation source provides a mechanism to sample a random wavelength from its configured spectrum. Most spectra are specified in tabular form and are thus sampled using inversion of the tabulated cumulative distribution. It is also possible, however, to hard-code a specialized sampling routine for analytically defined spectra \citep[as in, e.g.,][]{Carter1975}. It is important to properly sample all aspects of the spectrum, including narrow spectral features and wavelength ranges with lower luminosity. To this end, the sampling procedure employs specific biasing techniques described in Sect.~\ref{SamplingWavelengths}.

Finally, SKIRT~9 introduces the concept of a wavelength broadband with a given transmission curve. A broadband can be used to normalize a source luminosity or as part of a special wavelength grid for detecting fluxes in instruments. This feature is further described in Sect.~\ref{Bands}.

\subsection{Spatial distributions}
\label{SpatialDistributions}

One of the key SKIRT features is a suite of built-in geometries (e.g., an exponential disk) and decorators (e.g., spiral arms) designed to model spatial distributions \citep{Baes2015}. In previous versions of the code, a geometry could also support anisotropic emission. However, properly implementing this design for all available decorators turns out to be virtually impossible. We therefore decided to move anisotropic emission support away from the geometry concept and into a new category of sources (see Sect.~\ref{Sources}).

Likewise, as mentioned in Sect.~\ref{Challenges}, the existing geometry and decorator paradigm has been designed to model a density field normalized to unit mass and thus does not fit the new requirements for modeling velocity fields and for defining the direction and degree of grain alignment across the spatial domain. While it would be possible to design new geometry/decorator hierarchies for each of these new types of fields, we have chosen not to do so, at least for now. Instead, these new features are fully supported for sources and media that are imported from hydro-dynamical snapshots. Looking forward, these types of models are likely to be more pervasive than basic semi-analytical models. However, our generic design for sources and media does not preclude adding such built-in models in the future (see Sects.~\ref{Sources} and \ref{Media}).

SKIRT~9 includes a streamlined facility to enable importing complex snapshot data from text column files more easily. The new import module allows an input file to specify column names and the corresponding units as structured comments in the file header. The SKIRT configuration file can then refer to these names to specify column ordering. 

\subsection{Sources}
\label{Sources}

\begin{figure*}
  \centering
  \includegraphics[width=0.99\textwidth]{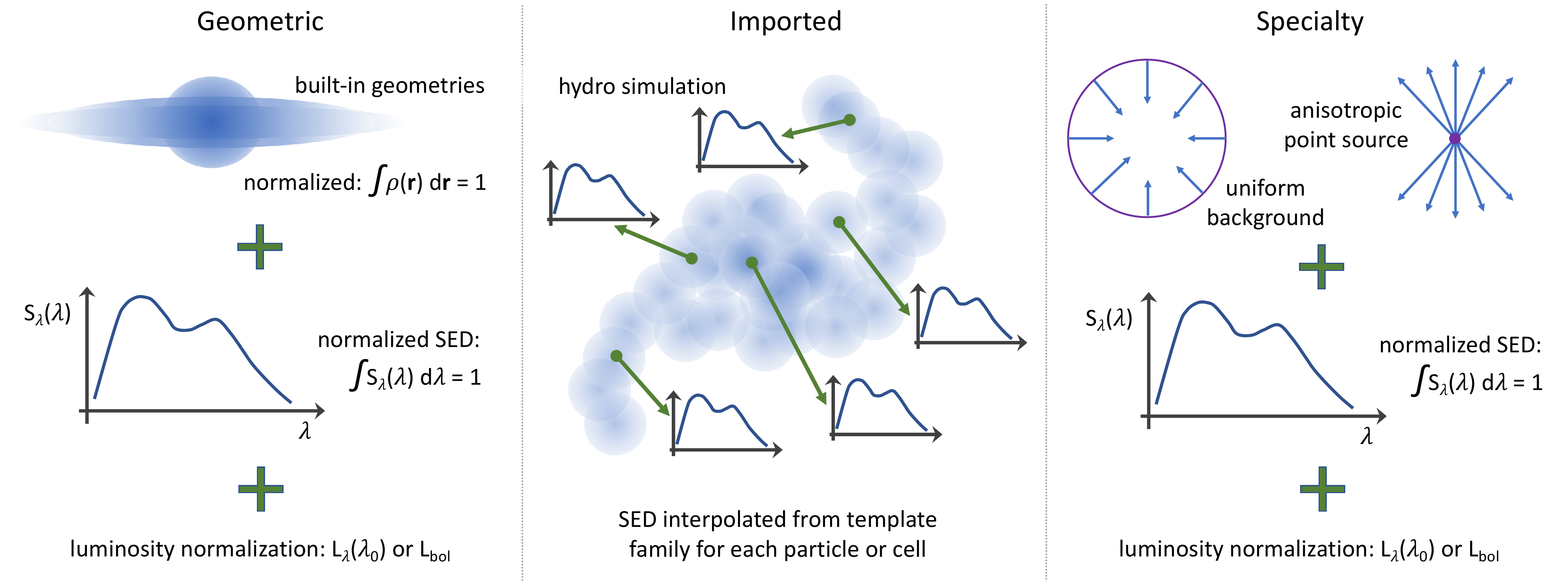}
  \caption{The three categories of primary sources in SKIRT~9. Geometric sources (left) combine built-in definitions for the spatial luminosity density distribution and for the spectral energy distribution (SED) with a specified luminosity normalization. Imported sources (middle) obtain both the spatial and spectral distribution from the particle or cell properties in the snapshot data generated by a hydro-dynamical simulation, given an SED template family. Specialty sources (right) include anisotropic point sources and sources of uniform background radiation. Sources of different types can be combined in a simulation.}
  \label{fig:sources9}
\end{figure*}

SKIRT~9 offers several types of primary radiation sources, including geometric, imported, and specialty sources (see Fig.~\ref{fig:sources9}). The configuration for a particular simulation can contain any number of sources and any mixture of source types. Regardless of its type, a source must define the following information at each point in the spatial domain:
\begin{itemize}
\item The spectral energy distribution (SED) of the emission averaged over the unit sphere.
\item Some normalization for the luminosity (for example, the luminosity at a given wavelength).
\item If the emission is not isotropic, the rest-frame angular distribution of the emission.
\item If the emission is polarized, the polarization state of the emission in each direction.
\item The velocity of the source relative to the model coordinate frame.
\end{itemize}

Furthermore, in order to emit a photon packet, a source must be able to sample a location from the spatial luminosity density distribution (see Sect.~\ref{DistributingPackets} for details), and then sample a wavelength from the SED at that location (see Sect.~\ref{SamplingWavelengths} for details).

For geometric sources, the spatial luminosity density distribution is defined by a combination of built-in geometries (e.g., exponential disk) and decorators (e.g., spiral arms) as described by \citet{Baes2015}. By definition, these sources have the same SED across their spatial domain (with a variable normalization according to the spatial luminosity density distribution), and their emission is isotropic and unpolarized.

Point sources and anisotropic sources such as those generating background radiation are now part of the new category of specialty sources. More specialty sources can be added at will. The implementation of a specialty source has maximum flexibility, as long as it conforms to the requirements listed above.

Imported sources implement sources defined by a set of particles or discretized on a structured or unstructured mesh, usually produced by a hydro-dynamical simulation. The emission spectrum for each particle or cell is selected from one of the built-in SED template families based on imported properties (for example, age and metallicity for a single stellar population). These sources offer several new features, such as the option to import the components of a velocity vector for each particle or cell. When more new features arrive, additional optional data columns can be easily added.

\subsection{Media}
\label{Media}

Similar to sources, SKIRT~9 offers two medium component types. Geometric medium components define a spatial mass density distribution through built-in geometries and decorators, and specify some normalization to determine the total mass of the component. Imported medium components load both spatial distribution and mass information from an input file using the new import module, plus an optional velocity vector for each particle or cell. Furthermore, each medium component has an associated material type, i.e. dust, electrons or gas.

All relevant properties for a particular material such as a mixture of dust grains with given composition and size distribution are bundled in a `material mix' object. A medium component usually has a single associated material mix, specifying identical material properties across its spatial domain. Imported medium components have the option to select a material mix from of a family of material mixes of the same type for each particle or cell.

Regardless of material type, each material mix must provide the material properties required for tracing photon packets through a material of this type, including the absorption and scattering cross sections and the scattering phase function properties (both as a function of wavelength). The SKIRT photon cycle supports both the Henyey-Greenstein phase function \citep{Henyey1941} and a custom, material mix-specific phase function that may depend on the polarization state of the incoming radiation as well as the scattering geometry for spherical grains.

Next to a material mix representing electrons, SKIRT includes a set of turn-key material mixes for various dust models that have been published in the literature \citep[e.g., ][]{Zubko2004, Draine2007, Jones2017}, and configurable dust mixes that allow specifying material compositions and grain size distributions for each component of the dust model. At the time of writing, the gas material type is provided but not actually implemented.

The configuration for a particular simulation can contain any number of medium components and any combination of material types and material mixes.

\subsection{Output}
\label{Output}

SKIRT conceptually generates two kinds of output. Most importantly, it produces synthetic observations obtained by detecting photon packets arriving at one of the instruments configured by the user. This output is written by each instrument at the end of the simulation run, and it is often the main simulation result. Secondly, it may be relevant to output the contents of internal data structures constructed in preparation for or during the simulation run. This includes diagnostics used to verify configuration and operation of the code and physical quantities that are computed by the simulation but cannot be `observed' from the outside.

The design of the instrument system has been streamlined in SKIRT~9. There are three types of instruments: the usual `distant' instruments using parallel projection, and the special purpose `all-sky' and `perspective' instruments. All instruments use the same `flux recorder' class to record the contributions of arriving photon packets into the appropriate flux density or surface brightness bins. This design avoids code duplication and ensures that each instrument offers all implemented options, such as tracking polarization (Stokes vector components) or separating fluxes on origin (direct or scattered, primary or secondary). The flux recorder, and thus all instruments, can also record and output information intended for calculating statistical properties such as the relative error $R$ on the results in each bin. Analysis of these statistical properties may help answering the question whether a sufficient number of photon packets was used in the simulation. This new feature is described in Sect.~\ref{Statistics}.

SKIRT~9 moves responsibility for producing the second type of output to a separate set of classes, called `probes', as opposed to embedding this functionality within the simulation code itself. From a user perspective, the SKIRT~8 `write' attributes scattered around the various simulation items are replaced by a list of probe objects living in a dedicated area of the run-time hierarchy (see Fig.~\ref{fig:configuration}). Examples include planar or linear cuts through the medium density, the medium temperature or the radiation field; a list of properties for each spatial cell; or the number of photon packets emitted from primary sources as a function of wavelength.

This updated design has several benefits. For the code developer and maintainer, it means that the simulation code is not cluttered with output code and there is an explicit interface between both types of code, resulting in improved data encapsulation. In fact, new probes can often be developed without changing the simulation code. For the user, it means that probe objects can take additional attributes to customize the output, for example, for requesting a cut through the medium density at some offset from the coordinate plane. Also, the configuration can contain multiple probes of the same kind, for example, for requesting a cut through the medium density at two different offsets from the coordinate plane.

In the current implementation, probes can get invoked at two points in the simulation: at the end of the setup phase, and at the end of the simulation run, just before the instruments generate output. Additional probe points may be added as they become relevant, for example after each step in the iterative process for calculating a self-consistent medium state (see Sect.~\ref{SelfConsistentState}).

\subsection{User interface}

\begin{figure*}
  \centering
  \includegraphics[width=0.99\textwidth]{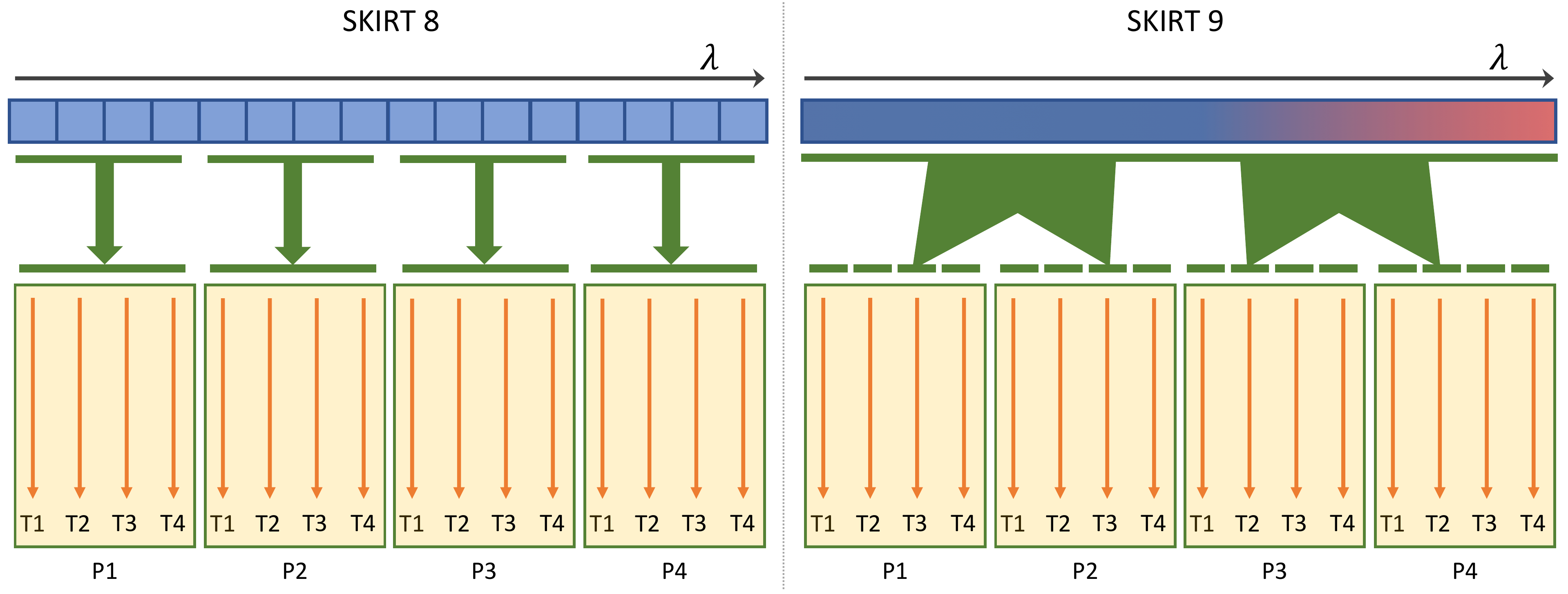}
  \caption{Parallel execution mechanisms before and after the transition discussed in Sect.~\ref{Parallel}. The green blocks (P1--P4) represent processes and the orange arrows (T1--T4) represent execution threads in each process. In SKIRT~8 (left), a predefined set of fixed wavelengths is assigned to each process. In SKIRT~9 (right), arbitrary wavelengths are sampled from the spectral distribution of the source (see Fig.~\ref{fig:wavelengths89}) for each execution thread in each process.}
  \label{fig:parallel89}
\end{figure*}

The SKIRT~9 configuration offers even more options than previous versions. This is inevitable because it supports more physics (see Sect.~\ref{Objectives}) and offers more flexibility (see, for example, the feature presentations in Sect.~\ref{Features}). This extra complexity might be confusing, especially to beginning or occasional users. To help alleviate this concern, we decided to improve the configuration process.

As presented by \citet{Camps2015a}, a SKIRT user can create a configuration file by responding to a series of questions in a command-line environment. The questions in this Q\&A session are fully driven by metadata embedded in the SKIRT code and depend on the user's responses to earlier questions in the session. In SKIRT~9, we substantially enhanced the capabilities to dynamically adjust the displayed options and the default values to choices made earlier in the configuration process. These earlier choices may represent various aspects of the configuration, such as for example the level of expertise selected by the user (basic, regular, advanced), the configured simulation mode (e.g., including secondary emission or not), or the level of spatial symmetry in the input model (1D, 2D or 3D). The metadata in the various C++ class definitions has subsequently been updated to hide options that are irrelevant or too advanced for a particular situation, providing appropriate default values in each case. See Sect.~\ref{Discovery} for more information.

These changes greatly simplify the configuration process, especially for a first-time user. The optional MakeUp utility takes this an important step further by offering a wizard-like graphical user interface for creating and adjusting configuration files, as described in Sect.~\ref{MakeUp}.

\subsection{Parallel execution}
\label{Parallel}

SKIRT~9 implements a hybrid parallelization mechanism similar to the SKIRT~8 \emph{task} parallelization mode \citep[see][]{Verstocken2017}. The total number of tasks at hand (e.g., the number of photon packets to be emitted) is divided into a number of chunks. These chunks are handed out dynamically to the various processes and/or threads in the simulation until all chunks are completed. The root process/thread keeps track of the number of completed chunks. This on-call scheme provides excellent load balancing.

When emitting a photon packet, the originating source and the packet's wavelength are, in principle, sampled from the appropriate probability distributions independently from any of the other photon packets (see Fig.~\ref{fig:parallel89}). As a result, each parallelization chunk potentially handles all sources and the complete wavelength range. In practice, photon packets from a particular source are usually emitted consecutively as described in Sect.~\ref{DistributingPackets}. This allows determining the emission spectrum for the source only once without the need for long-term storage (just one spectrum needs to be stored for each execution thread). This is especially important when this calculation is expensive, such as when interpolating a stellar population spectrum from a multi-dimensional family of SED templates \citep[e.g.,][]{Maraston1998, Castelli2003, Groves2008}, or when handling the emission by stochastically heated dust grains in a given spatial cell \citep{Camps2015b}.

Even with this optimized photon packet emission ordering, each parallelization chunk still potentially covers the complete wavelength range of the simulation. Furthermore, the effects of kinematics can cause the wavelength of a photon packet to change during its life cycle. As a result, it is impossible (or at least far from trivial) for SKIRT~9 to implement the SKIRT~8 \emph{data} parallelization mode, which assigns a specific set of wavelengths to each process \citep[see][]{Verstocken2017}. Fortunately, when configured properly, SKIRT~9 is substantially more memory efficient than SKIRT~8 (see Sect.~\ref{Memory}), eliminating the need for a data-distributed mode in many cases. Still, we would like to revisit this issue in a future version. One might envision a scheme that assigns a wavelength range to each process and communicates photon packets that drift outside of the assigned range to the appropriate process. Alternatively, one might consider a full spatial domain decomposition scheme, where photon packets are communicated between processes as they cross the borders of each local domain \citep[as in, e.g.,][]{Harries2019}. It remains to be evaluated how the communication overhead inherent in such schemes would impact the total run-time of the simulation.

\begin{table*}
  \caption{The repositories in the SKIRT organization on GitHub (https://github.com/SKIRT).}
  \label{tab:repositories}
  \centering
  \begin{tabular}{lll}
  \hline
  Repository & Description & Status \\
  \hline
  SKIRT7 & SKIRT 7 C++ source code & Deprecated \\
  SKIRT8 & SKIRT 8 C++ source code & Maintenance mode \\
  SKIRT9 & SKIRT 9 C++ source code & Current development \\
  PTS & Python toolkit for SKIRT 7 and 8 & Maintenance mode \\
  PTS9 & Python toolkit for SKIRT 9 & Current development \\
  Web8 & SKIRT 8 and PTS 8 documentation & Maintenance \\
  Web9 & SKIRT 9 and PTS 9 docs + project pages & Current development \\
  \hline
  \end{tabular} 
\end{table*}

\subsection{Python toolkit}
\label{PythonToolkit}

The Python Toolkit for SKIRT (PTS) was originally developed as a set of Python modules offering functionality for working with SKIRT~7 and SKIRT~8. Over the years, it has far outgrown its original scope, now including sub-packages related to specific projects, often with a limited or no connection to SKIRT \citep[e.g.,][]{Clark2018, Decleir2019, Verstocken2020, Nersesian2020}. The size of the code base raises significant concerns about maintenance and support. While several of the sub-packages might make sense as stand-alone projects or as packages affiliated with other open-source efforts, the scope of the combined PTS project is too wide and too varied. There are also many dependencies between the various sub-packages, so that carving out any of them would be a nontrivial undertaking. Finally, for historical reasons, PTS is written in Python version 2, while modern code should use Python 3.

For these reasons, we decided to re-implement the core SKIRT-related PTS functionality as a new version, called PTS~9, in a fresh code repository. PTS~9 is written in Python 3.7 and provides modules to work with SKIRT~9, without support for earlier versions. The previous PTS repository, with all its features, including support for earlier SKIRT versions, remains available in maintenance mode. Any interested party is welcome to re-purpose any portion of the code and make it available as a standalone package or as part of some other effort.

PTS 9 includes functions for interfacing with SKIRT (handle configuration files, perform a simulation, write input files and load output files), visualizing SKIRT output (image frames, SEDs, density cuts, temperature cuts, polarization maps, and more), and supporting the SKIRT development process (run test cases and benchmarks). These functions can be accessed from the command line, from interactive notebooks, and through regular Python function calls. The PTS~9 source code is publicly available on GitHub\footnote{https://github.com/SKIRT/PTS9} and the documentation can be found on the SKIRT web site.\footnote{http://www.skirt.ugent.be}

\subsection{Deployment}

As indicated in Sect.~\ref{Challenges}, we intend to actively establish a collaborative environment for the SKIRT project. All previous and current SKIRT project repositories are publicly available as part of the SKIRT organization on GitHub; see Table~\ref{tab:repositories} for a list. There are no longer any private repositories. Maintaining just a single copy of each repository greatly simplifies the workflow for handling contributions, and ensures that everyone has access to the most recent version of the code. Next to the SKIRT and PTS source code, the source text for the SKIRT web site is also placed in a public repository, allowing contributions to the documentation (including user guide, tutorials, and more) through a workflow similar to that for source code.

Because the repositories are public, any GitHub user has read access and can post an issue or send a pull request. The SKIRT organization uses the fork and pull workflow model. Anyone can fork a repository, push changes to their personal fork, and initiate a pull request. The changes can be pulled into the source repository by a core team member, possibly after discussion and/or being adjusted in one or more iterations.

Usage questions, bug reports and feature requests are managed through the issues system offered by GitHub. Anyone can post an issue, and anyone can respond. Core team members make sure that issues are addressed timely, label and assign issues appropriately, and eventually close issues as needed.

A user with a specific interest in SKIRT can become a member of the SKIRT Contributor team, granting them access to SKIRT-related broadcast notifications and team discussions. Notifications about important events (e.g., a new major feature becoming available, or an upcoming SKIRT user group meeting) can be broadcast to all contributors through a GitHub team page. The team page also allows discussions that are not directly related to a particular issue (e.g., about a particular SKIRT use case or an upcoming conference).

Starting with SKIRT~9, the GitHub workflow automatically performs a `continuous integration' test with each commit to the master branch, and with each pull request targeted towards that same branch. If the test fails, the commit or pull request is refused. This is accomplished through the Travis CI service\footnote{travis-ci.com}, which is free of charge for open source projects. Prompted by a commit or pull request, Travis CI instantiates one or more virtual computer systems with predefined software configurations. Each of these virtual systems is instructed to pull the SKIRT code from the appropriate branch, compile and build it, possibly perform additional tests, and report a success/failure state for viewing within the regular GitHub web interface. At the time of writing, Travis CI builds the SKIRT code on three common operating system/compiler combinations. In the future, we can consider also performing (some of) the functional tests discussed in Sect.~\ref{FunctionalTests}.


\section{Design of selected features}
\label{Features}

\subsection{Distributing photon packets over sources}
\label{DistributingPackets}

A single source system object represents the primary source of radiation in a SKIRT simulation, consisting of the superposition of one or more sources. Each source provides a complete description of its emitted radiation, including the spatial luminosity density distribution. One key task of the source system is to distribute photon packet launches across the sources. In principle, this would be achieved by randomly selecting a source for each launch by sampling from an appropriate probability distribution \citep[see, e.g.,][]{Baes2015}. However, for some sources, a deterministic approach allows significant performance optimizations. Because the number of photon packets usually is much larger than the number of sources, a deterministic approach can be considered to be equivalent to the randomized procedure.

The idea is to iterate through the sources and launch consecutive photon packets from each. A source consisting of many sub-sources (such as imported particles or cells) can then use a similar approach, iterating over these sub-sources. The implementation can construct and cache relevant data structures (such as a cumulative spectral distribution) for each sub-source, and release the information as soon as the iteration moves on to the next sub-source. Because photon packets are often launched in parallel, these data structures must be allocated in thread-local storage, but that is only a minor complication.

As a first step, the total number $N$ of photon packets to be launched is passed to the source system in serial mode. Subsequently, packets are launched in parallel mode, and each packet is labeled with a unique index in the range $0,...,N-1$. Parallel execution threads handle chunks of photon packets with consecutive indices within a given subset of this range (see Sect.~\ref{Parallel}).

To achieve the goals described above, when it is passed the number $N$, the source system maps consecutive index ranges to each of the sources being held. This mapping is also passed on to each source, so that it can (but doesn't have to) implement a similar approach for its sub-sources. The number of photon packets allocated to each source is determined according to a composite biasing scheme \citep{Baes2016} as follows:
\begin{linenomath}\begin{equation}
N_s = \left[ (1-\xi) \frac{w_s L_s}{\sum_s w_s L_s} + \xi \frac{w_s}{\sum_s w_s} \right] N
\end{equation}\end{linenomath}
where $N$ is the total number of photon packets to be launched, $N_s$ is the number of photon packets to be launched by source $s$, $L_s$ is the bolometric luminosity of source $s$, $w_s$ is the \emph{source weight} for source $s$, $\xi$ is the \emph{source bias} of the source system, and the sums range over all sources in the source system.

By default, the source weights are set to $w_s=1,\forall s$ and the source bias is $\xi=\nicefrac{1}{2}$, which means that one half of the photon packets is distributed proportionally to the luminosity of the sources, and the other half is distributed equally over the sources. Changing the source weights $w_s$ allow a user to assign more importance to particular sources, and the bias factor $\xi$ can be adjusted to swing between the proportional and linear allocation schemes.

Imported sources (see Sect.~\ref{Sources}) in turn map their assigned range of photon packet indices to each of the sub-sources (imported particles or cells) using a similar biasing scheme:
\begin{linenomath}\begin{equation}
\label{eq:subsources}
N_m = \left[ (1-\xi) \frac{L_m}{L} + \xi \frac{1}{M} \right] N_s
\end{equation}\end{linenomath}
where $N_s$ is the number of photon packets to be launched by this source, $N_m$ is the number of photon packets to be launched by sub-source $m$, $L_m$ is the luminosity of sub-source $m$, $L$ is the total luminosity of this source, $M$ is the number of sub-sources, and $\xi$ is the source bias of the source system (i.e.\ the same value as above).

The secondary source system, which handles the emission by the medium components in the simulation, uses the scheme described by Eq.~(\ref{eq:subsources}) to distribute photon packets over the cells in the spatial grid discretizing the simulation domain. The index $m$ now ranges over the cells, and the value of the bias fraction $\xi$ can be configured separately for primary and secondary sources.

In all of the allocation schemes discussed in this section, the luminosity weight assigned to each photon packet is adjusted to compensate for the various biasing factors. 

\subsection{Sampling wavelengths}
\label{SamplingWavelengths}

\begin{figure}
  \centering
  \includegraphics[width=0.99\columnwidth]{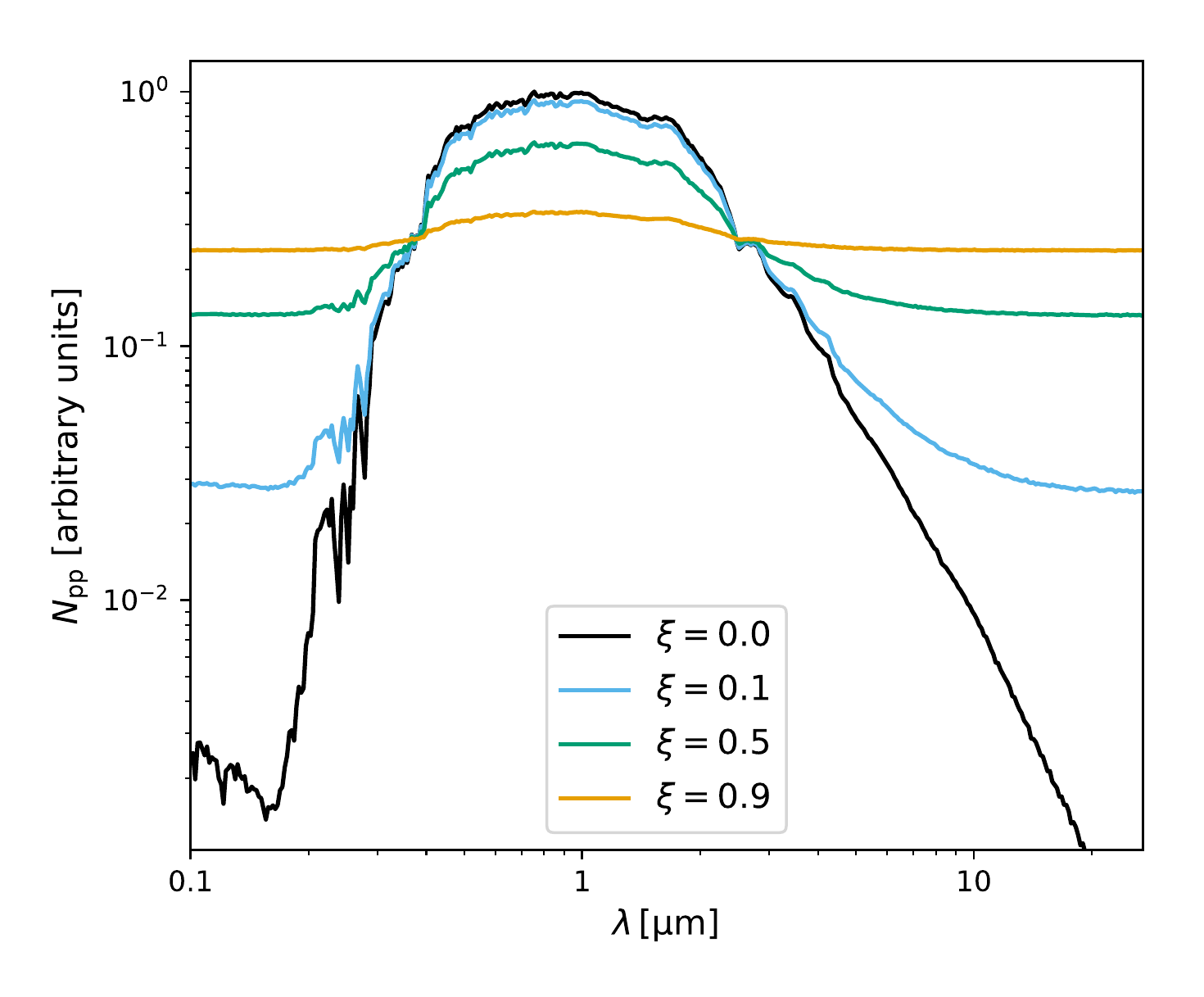}
  \caption{The number of photon packets launched as a function of wavelength according to Eq.~(\ref{eq:wavebias}) for a source with a typical single stellar population spectrum $s(\lambda)$, using a logarithmic bias distribution $b(\lambda)\propto 1/\lambda$ and bias fractions $\xi$ ranging from no bias ($\xi=0$), i.e.\ following the source spectrum, to very strong bias.}
  \label{fig:wavebias}
\end{figure}

When a photon packet is being emitted from a primary or secondary source, its rest-frame wavelength is, in principle, randomly sampled from the source's spectral energy distribution. In practice, wavelengths for new photon packets are sampled from a linear combination of the intrinsic spectral distribution of the source $s(\lambda)$ and a bias wavelength distribution $b(\lambda)$. Given these distributions and a bias fraction $\xi$, the composite distribution $q(\lambda)$ is
\begin{linenomath}\begin{equation}
\label{eq:wavebias}
q(\lambda) = (1-\xi) s(\lambda) + \xi b(\lambda)
\end{equation}\end{linenomath}
and the corresponding biasing weight factor, applied to the photon packet's luminosity, becomes \citep[see][]{Baes2016}
\begin{linenomath}\begin{equation}
w(\lambda) = \frac{s(\lambda)}{q(\lambda)} = \frac{s(\lambda)}{(1-\xi) s(\lambda) + \xi b(\lambda)}
\end{equation}\end{linenomath}

Both the bias fraction $\xi$ and the bias distribution $b(\lambda)$ can be configured by the user. By default, $\xi=\nicefrac{1}{2}$, so that half of the photon packet wavelengths are sampled from each of the distributions. The default bias distribution is one for which the logarithm of the wavelength is distributed uniformly, which corresponds to $b(\lambda)\propto 1/\lambda$ with proper normalization over the wavelength range of the source.

As illustrated in Fig.~\ref{fig:wavebias}, the default scheme ($\xi=0.5$) ensures that the low-luminosity tails of a typical spectrum are properly sampled, while still favoring the higher-luminosity areas. Even narrow spectral features are properly sampled because half of the wavelengths are selected from the source SED at full spectral resolution (see also the discussion in Sect.~\ref{Comparison89}). Lowering the bias fraction (e.g., $\xi=0.1$) focuses more photon packets into high-luminosity areas because the composite distribution more closely follows the source spectrum. Vice versa, a bias fraction close to unity (e.g., $\xi=0.9$) causes the source spectrum to be essentially ignored for the purpose of wavelength sampling.

The default bias distribution is usually appropriate for wavelength ranges spanning multiple decades, where one aims for a constant spectral resolution $R=\lambda/\Delta\lambda$ over the entire range (modulated with the source spectrum as per the bias fraction $\xi$). For narrow wavelength ranges, perhaps corresponding to a particular spectrograph or spanning a given emission line, a linear distribution of the photon packet wavelengths might be more appropriate. To this end, SKIRT~9 offers a built-in uniform wavelength bias distribution in addition to the logarithmic distribution. Users can also load a custom distribution from file for maximum flexibility. For example, one might want to strongly favor a wavelength interval of special interest even if the model's sources are not particularly luminous in that interval.

\subsection{Kinematics}
\label{Kinematics}

As discussed in Sect.~\ref{Wavelengths}, a SKIRT~9 photon packet remembers its weight $w$, a measure for the number of photons (per unit of time) in the packet, and its wavelength $\lambda$, valid for all of those photons. The luminosity $L$ carried by the packet is thus proportional to $w$ and inversely proportional to $\lambda$. To avoid multiplying and dividing by a constant factor, the weight $w$ is stored in units so that simply $L=w/\lambda$.

Given the photon packet's current direction $\bf{\hat{k}}$ (a unit vector) and the velocity of a medium $\bf{v}_\mathrm{m}$ relative to the model coordinate frame, the Doppler-shifted wavelength $\lambda_\mathrm{m}$ perceived by the medium is easily obtained by
\begin{linenomath}\begin{equation}
\left. \lambda_\mathrm{m} = \lambda \middle/ \left(1 - \frac{\bf{\hat{k}} \cdot \bf{v}_\mathrm{m}}{c} \right) \right.
\end{equation}\end{linenomath}
where $c$ is the speed of light in vacuum and we assume non-relativistic medium velocities. The perceived wavelength $\lambda_\mathrm{m}$ is used during the photon life cycle to obtain extinction cross sections in spatial cells along the photon packet path, to calculate the characteristics of scattering events, and to determine the appropriate wavelength bin for recording the photon packet's contribution to the radiation field (see Sect.~\ref{RecordingField}). In the latter case, the luminosity perceived by the medium is calculated as $L_\mathrm{m}=w/\lambda_\mathrm{m}$, properly taking into account the Doppler shift here as well.

Conversely, the Doppler-shifted wavelength that should be assigned to a photon packet when it is emitted into direction $\bf{\hat{k}}$ from a source moving with velocity $\bf{v}_\mathrm{s}$ relative to the model coordinate frame can be written as 
\begin{linenomath}\begin{equation}
\lambda = \lambda_\mathrm{s} \left(1 - \frac{\bf{\hat{k}} \cdot {\bf{v}}_\mathrm{s}}{c} \right)
\label{eq:emissionshift}
\end{equation}\end{linenomath}
where $\lambda_\mathrm{s}$ is the wavelength emitted in the rest-frame of the source. This Doppler-shift is applied when a photon packet is emitted from a moving primary source or from the moving medium in a spatial cell, and after a photon packet scatters off a moving medium. In the latter case, the medium velocity acts as the source velocity in Eq.~(\ref{eq:emissionshift}).

\subsection{Recording the radiation field}
\label{RecordingField}

To enable the calculation of the secondary emission by the medium components in a model, a panchromatic radiative transfer simulation needs to determine the radiation field in every cell of the simulation's spatial grid. This is done by tracking the relevant contribution, as a function of wavelength, for each photon packet moving through a cell. Earlier versions of SKIRT stored the portion of a photon packet's luminosity absorbed by the dust in the cell being crossed (accumulated in the appropriate wavelength bin). However, if the simulation includes a gas medium next to dust, this dust-specific information becomes ambiguous or even meaningless. In SKIRT~9, therefore, we instead accumulate a quantity that does not depend on the medium present in the cell, using a method adapted from \citet{Lucy1999}. As a side benefit, this means that the radiation field can be tracked even for empty cells.

For each spatial cell $n$ along a photon packet's path, the function responsible for recording the radiation field first determines the wavelength bin $\ell$ corresponding to the photon packet's perceived wavelength $\lambda_n$ in that cell (see Sect.~\ref{Kinematics}). To the contents of that wavelength/cell bin, the function adds the product of the mean luminosity $\left<L\right>_n$ carried by the photon packet along the path segment in the cell and the length $(\Delta s)_n$ of that segment, or 
\begin{linenomath}\begin{equation}
(L\Delta s)_{\ell,n} \triangleq \left<L\right>_n\,(\Delta s)_n \,.
\end{equation}\end{linenomath}
To calculate the mean luminosity $\left<L\right>_n$ in this equation, we assume exponential behavior of the extinction along the path segment in the cell, corresponding to our overall assumption in SKIRT that the optical properties and the density of the medium are constant within each cell.\footnote{Dichroic media do not exhibit pure exponential extinction even for constant density, but we assume that the cells are sufficiently small for the discrepancies to be insignificant; see also Sect.~\ref{Polarization}.} Given the photon packet luminosity $L_n$ perceived by the cell (see Sect.~\ref{Kinematics}), the cumulative optical depths $\tau_{n-1}$ and $\tau_n$ along the path at the start and end of the segment in the cell, and writing $k_n$ for the opacity of the medium in the cell, all determined at the perceived wavelength $\lambda_n$, we obtain,
\begin{linenomath}\begin{equation}
\begin{split}
\left<L\right>_n & = \frac{\int_{(\Delta s)_n} L(s)\,\mathrm{d}s}{\int_{(\Delta s)_n} \mathrm{d}s} \\
& = \frac{1}{(\Delta s)_n} \int_0^{(\Delta s)_n} L_n\, \mathrm{e}^{-\tau_{n-1}} \mathrm{e}^{-k_n s} \,\mathrm{d}s \\
& = \frac{1}{(\Delta s)_n} L_n\, \mathrm{e}^{-\tau_{n-1}} \left(\frac{1-\mathrm{e}^{-k_n s}}{k_n}\right) \\
& = L_n\;\frac{\mathrm{e}^{-\tau_{n-1}} - \mathrm{e}^{-\tau_n}}{\tau_n - \tau_{n-1}} \\
& = L_n\,\mathrm{M}_\mathrm{ln} \left(\mathrm{e}^{-\tau_{n-1}}, \mathrm{e}^{-\tau_n}\right) 
\end{split}
\end{equation}\end{linenomath}
where $\mathrm{M}_\mathrm{ln}()$ denotes the logarithmic mean \citep{Carlson1972}.

Once this information has been accumulated for all photon packets launched during a given phase of the simulation, the mean intensity of the radiation field in each spatial/wavelength bin can be calculated using
\begin{linenomath}\begin{equation}
(J_\lambda)_{\ell,m} = \frac{ (L\Delta s)_{\ell,m} }{4\pi\,V_m\,(\Delta \lambda)_\ell} 
\end{equation}\end{linenomath}
where $(\Delta \lambda)_\ell$ is the wavelength bin width, $m$ is the spatial cell index, $V_m$ is the volume of the cell, and $(L\Delta s)_{\ell,m}$ has been accumulated over all photon packets contributing to the bin. 

The bolometric luminosity absorbed by a given medium in the spatial cell with index $m$ can similarly be calculated using
\begin{linenomath}\begin{equation}
L^\mathrm{abs}_{\mathrm{bol},m} = \sum_\ell k^\mathrm{abs}_{\ell,m} \,(L\Delta s)_{\ell,m}
\end{equation}\end{linenomath}
where $\ell$ runs over the wavelengths in the simulation's radiation field wavelength grid, and $k^\mathrm{abs}_{\ell,m}$ is the wavelength-dependent absorption opacity of the medium in the cell. 

\begin{figure*}
\centering
\begin{minipage}[c]{0.93\textwidth}
\begin{lstlisting}[style=cpp]
class ClumpyGeometryDecorator : public Geometry
{
    ITEM_CONCRETE(ClumpyGeometryDecorator, Geometry, "a decorator that adds clumpiness to any geometry")
        ATTRIBUTE_TYPE_INSERT(ClumpyGeometryDecorator, "Dimension3")

    PROPERTY_ITEM(geometry, Geometry, "the geometry to be made clumpy")

    PROPERTY_DOUBLE(clumpFraction, "the fraction of the mass locked up in clumps")
        ATTRIBUTE_MIN_VALUE(clumpFraction, "[0")
        ATTRIBUTE_MAX_VALUE(clumpFraction, "1]")

    PROPERTY_INT(numClumps, "the total number of clumps")
        ATTRIBUTE_MIN_VALUE(numClumps, "1")

    PROPERTY_DOUBLE(clumpRadius, "the scale radius of a single clump")
        ATTRIBUTE_QUANTITY(clumpRadius, "length")
        ATTRIBUTE_MIN_VALUE(clumpRadius, "]0")

    PROPERTY_BOOL(cutoffClumps, "cut off clumps at the boundary of the underlying geometry")
        ATTRIBUTE_DEFAULT_VALUE(cutoffClumps, "false")
        ATTRIBUTE_DISPLAYED_IF(cutoffClumps, "Level2")

    PROPERTY_ITEM(smoothingKernel, SmoothingKernel, "the smoothing kernel that describes the density of a clump")
        ATTRIBUTE_DEFAULT_VALUE(smoothingKernel, "CubicSplineSmoothingKernel")
        ATTRIBUTE_DISPLAYED_IF(smoothingKernel, "Level2")

    ITEM_END()

protected:
    void setupSelfAfter() override;

public:
    double density(Position bfr) const override;
    Position generatePosition() const override;

    ...
};
\end{lstlisting}
\end{minipage}
\caption{The class declaration for a simulation item used to introduce random clumps into an otherwise smooth density distribution. This code is the SKIRT~9 equivalent of the earlier version shown in figure 15 of \citet{Camps2015a}. The macro invocations on lines 3-27 replace the Qt-based metadata specifications in the earlier version, with additional functionality as described in Sect.~\ref{Discovery}.}
\label{fig:metadatageom}
\end{figure*}

\begin{figure*}
\centering
\begin{minipage}[c]{0.93\textwidth}
\begin{lstlisting}[style=cpp]
class Sphere1DSpatialGrid : public SphereSpatialGrid
{
    ITEM_CONCRETE(Sphere1DSpatialGrid, SphereSpatialGrid, "a spherically symmetric spatial grid")
        ATTRIBUTE_TYPE_ALLOWED_IF(Sphere1DSpatialGrid, "!Dimension2&!Dimension3")

    PROPERTY_ITEM(meshRadial, Mesh, "the bin distribution in the radial direction")
        ATTRIBUTE_DEFAULT_VALUE(meshRadial, "LinMesh")

    ITEM_END()

    ...
};
\end{lstlisting}
\end{minipage}
\caption{The class declaration for a simulation item used to configure a spherically symmetric spatial grid. The macro invocation on line 4 ensures that this grid can only be selected if the input model has the appropriate symmetry.}
\label{fig:metadatagrid}
\end{figure*}

\subsection{Wavelength broadbands}
\label{Bands}

A \emph{band} object in SKIRT~9 represents the transmission curve of a particular observational filter as a function of wavelength. Key operations offered by all band objects include obtaining the transmission at a given wavelength and calculating the mean specific luminosity for a given SED after convolution with the transmission curve.

SKIRT~9 offers a set of built-in band objects for standard filters, such as the Johnson filters, and common observatories, such as GALEX \citep{Morrissey2007}, SDSS \citep{Doi2010}, and Herschel \citep{Poglitsch2010, Griffin2010}. For example, all broadbands listed in table 4 of \citet{Camps2018a} are included. Other bands can be loaded from file, given a tabulated transmission curve.

A band object can be used to normalize the luminosity of a source to a given mean specific luminosity for the band, which often corresponds more precisely to an observed quantity than specifying a specific luminosity at a particular wavelength. More interestingly, perhaps, it is also possible to equip an instrument with a `wavelength grid' built from a list of (possibly overlapping) bands. In that case, each band represents a separate bin of the instrument. When a photon packet arrives, its contribution is multiplied by the transmission at the packet's wavelength for each band before being accumulated in the corresponding bin. This amounts to `on-the-fly' convolution of the detected flux with the transmission curve of each band.

A radiative transfer simulation is often performed with the aim of comparing its results with observations. In that case, using a band wavelength grid produces directly comparable output. The alternative is to run the simulation using a regular wavelength grid with fairly narrow bins, and perform the convolution after the fact. For proper results, the instrument wavelength grid must resolve all spectral features of the sources, including emission or absorption lines which may be Doppler shifted because of kinematic effects. This may require a large number of bins, with correspondingly large memory requirements (see Sect.~\ref{Memory}).

\subsection{Polarization by aligned spheroidal grains}
\label{Polarization}

SKIRT~9 inherits support for polarization caused by scattering off spherical dust grains from earlier versions \citep{Peest2017} and adds the option to specify polarized emission for primary sources (see Sect.~\ref{Sources}). Over time, we aim to also model the polarization effects of non-spherical dust grains that are partially aligned by magnetic fields (see Sect.~\ref{Objectives}). Apart from the ability to specify the grain alignment degree and direction in the input model, as discussed in Sect.~\ref{SpatialDistributions}, this will require extensions in several areas of the code \citep[see, e.g.,][]{Reissl2014, Reissl2016}. 

For example, defining the geometry for scattering by a spheroidal grain requires two additional angles to specify the directions of the incoming and outgoing photon packet relative to the grain's symmetry axis \citep[e.g.,][]{Wolf2002}. The optical material properties that govern the transformation of the polarization state as a result of the scattering event depend on these additional angles, in practice requiring a much larger set of tabulated data. Similarly, the polarization state of radiation emitted from aligned spheroidal grains depends on the outgoing direction relative to the grain's symmetry axis, again requiring the appropriate tabulated optical material properties.

Perhaps more fundamentally, a medium with aligned dust grains is dichroic. This means that the polarization state of radiation changes as it moves through the medium and that the extinction by the medium depends on the local polarization state of the radiation. As a result, the calculations to move a photon packet along its path become significantly more complicated \citep[see, e.g.,][]{Mishchenko1991, Baes2019}.

\subsection{Self-consistent medium state calculations}
\label{SelfConsistentState}

For models where self-absorption of thermal dust emission may be significant, SKIRT iterates over the dust emission phase until the state of the dust medium has converged. As indicated in Sect.~\ref{Objectives}, some of the physical processes that may be added to SKIRT in the future will require a similar self-consistent calculation of the medium state in equilibrium with the radiation field. Examples include determining the ionization state and/or level populations of hydrogen \citep{Osterbrock1989} and modeling dust destruction \citep{Jones2004} near energetic sources. We plan to add these iterations as they become needed.

\subsection{Metadata-driven user interface}
\label{Discovery}

As described by \citet{Camps2015a}, the core run-time data structure for a SKIRT simulation consists of a tree-like hierarchy of \emph{simulation items}, i.e.\ instances of a \cpp{SimulationItem} subclass. This data structure fully defines the configuration of the simulation, including the input model, the wavelength regime, the requested outputs, and all related options. The C++ class definitions for the \cpp{SimulationItem} subclasses provide the metadata required to drive a user-friendly command-line question-and-answer (Q\&A) session for creating an XML-formatted configuration file from which the run-time hierarchy can be constructed. As a result, the Q\&A and the configuration file structure automatically adapt to changes in and additions to the SKIRT functionality.

In the version of SKIRT described by \citet{Camps2015a}, the implementation of this metadata-driven user interface was based on the Qt\footnote{https://www.qt.io} cross-platform run-time environment. In SKIRT~9, the simulation code and the command-line Q\&A no longer depend on Qt, facilitating installation on remote servers and multi-node computing systems. Fig.~\ref{fig:metadatageom} shows the SKIRT~9 class declaration for a typical simulation item, namely a geometry decorator used to introduce random clumps into the spatial density distribution defined by another, arbitrary geometry. The macros invoked on lines 3-27 are defined in a header file that is included for every simulation item. They replace and augment the Qt-based implementation in earlier SKIRT versions by a regular C++14 implementation\footnote{Purists will argue that the macro pre-processor facility is not really a part of the C++ language. While we agree with this viewpoint in principle, we also believe that sometimes one needs to be pragmatic. SKIRT~9 includes just a single, well-documented header file defining the macros that achieve our goals in combination with a much larger code base of proper C++ constructs.}, eliminating the Qt dependency from the core SKIRT code.

The \cpp{ITEM} macro (line 3) ensures that the class can be registered to the module driving the user interface and specifies a human-readable title. The \cpp{PROPERTY} macros specify the configurable properties for the simulation item. These properties can have various data types including scalar values (lines 6, 8 and 15) and aggregation types that include more items in the hierarchy (lines 6 and 23). The interspersed \cpp{ATTRIBUTE} macros specify extra information about the preceding item or property.

Each \cpp{PROPERTY} macro generates code in the class definition for various aspects of the specified property:
\begin{itemize}
\item A mechanism to register the relevant metadata including the property type and title to the user interface module.
\item A private data member declaration for the property; this data member can be accessed in the class implementation.
\item A public getter for the property value; this getter can be freely used inside and outside of the class implementation.
\item A private setter with an undocumented name used by the user interface module to initialize the property value when constructing a new simulation item hierarchy from a configuration file.
\end{itemize}
Centralizing these responsibilities in the macro definitions avoids error-prone repetitive code, for example for defining data members and property access functions in the class header.

The \cpp{ATTRIBUTE} macro capabilities have grown as well. For example, the minimum and maximum values for a \cpp{double} property (lines 9-10 and 17) can have square brackets to indicate whether or not the limiting value is included in the allowed range. As a result, many simulation item classes, including the one shown in the figure, no longer need the \cpp{setUpSelfBefore()} function, which was often present in previous SKIRT versions just to test these limiting cases.

The \cpp{ATTRIBUTE} macro invocations on lines 21 and 25 specify that the corresponding property should be displayed only if the user has established experience level 2 or higher. The condition specified for this type of \cpp{ATTRIBUTE} macros can be an arbitrary Boolean expression using `variables' defined earlier in the configuration process. For example, the \cpp{INSERT} macro invocation on line 4 in Fig.~\ref{fig:metadatageom} inserts the variable \cpp{Dimension3} with a value of \cpp{true} whenever a clumpy geometry decorator is part of the configuration. This information can then be tested elsewhere, such as by the macro invocation on line 4 of Fig.~\ref{fig:metadatagrid}. In this case, the result is that the spherically symmetric spatial grid will not be offered as a choice if the configuration includes a clumpy geometry decorator. Assuming that all geometry classes provide the correct \cpp{INSERT} specifications, this will hold for any geometry that does not conform to a one-dimensional symmetry.

\begin{figure}
  \centering
  \includegraphics[width=0.99\columnwidth]{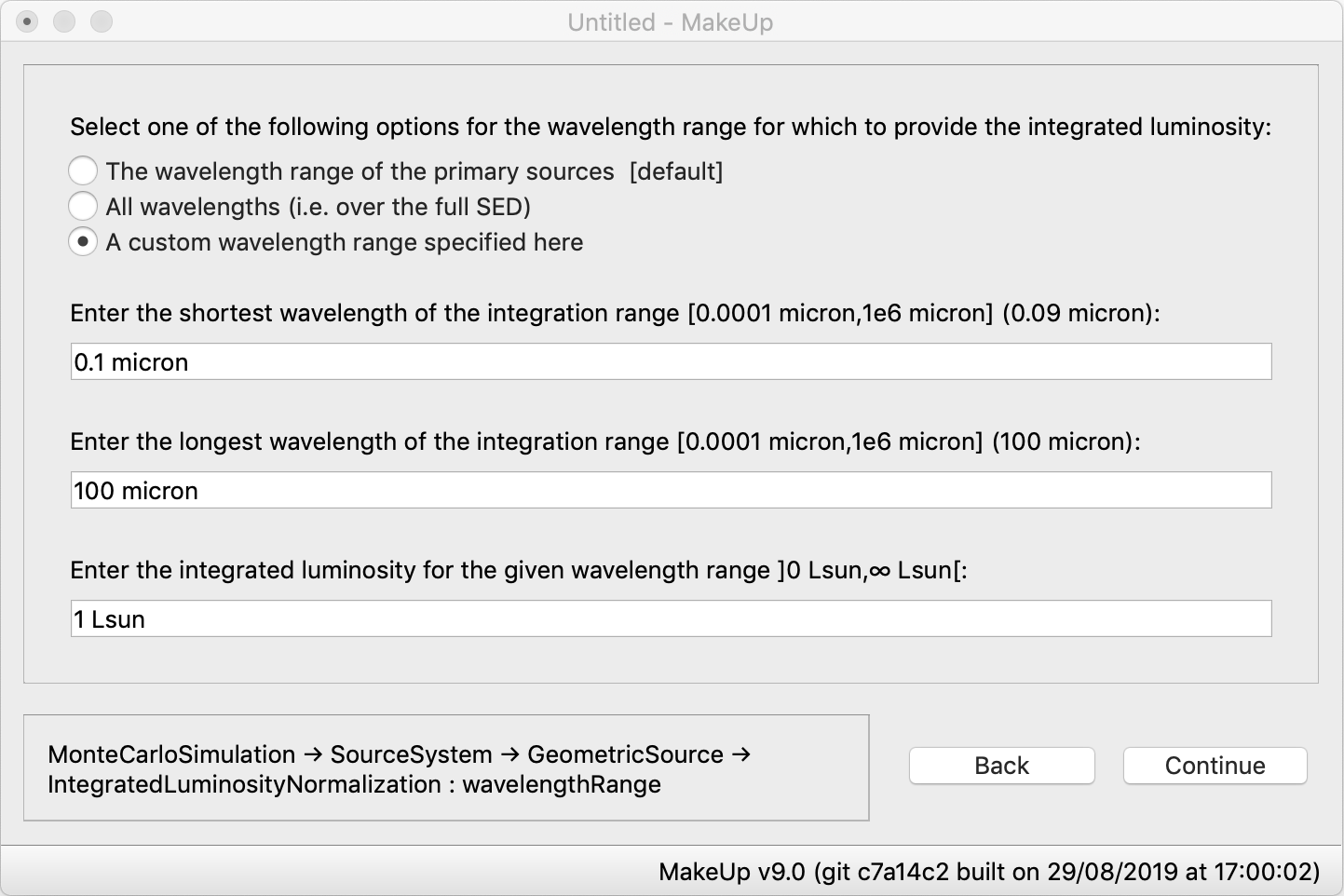}
  \caption{The graphical user interface of the MakeUp wizard while configuring the integrated luminosity for a particular radiation source.}
  \label{fig:makeupcustom}
\end{figure}

\subsection{The graphical configuration wizard MakeUp}
\label{MakeUp}

While the SKIRT~9 command-line Q\&A facility and the simulation code itself no longer depend on the Qt framework, we used Qt to develop a cross-platform graphical user interface for creating and adjusting SKIRT configuration files. Because this utility, called MakeUp, is distinct from the SKIRT code itself, the configuration process and the simulation itself can run on different computer systems, for example a desktop and a remote server, respectively. In this case, only the desktop computer needs to have the Qt environment installed.

MakeUp offers a wizard-like interface that is fully driven by the metadata used for the command-line Q\&A facility, as discussed in Sect.~\ref{Discovery}. In fact, as part of the SKIRT build process, all of the relevant metadata is bundled into a single file, called a \emph{schema}, which can then be loaded by MakeUp when it is launched.

Fig.~\ref{fig:makeupcustom} shows a screenshot of the MakeUp interface while configuring the integrated luminosity for a particular radiation source. The set of radio buttons at the top corresponds to an enumeration property in the metadata definition of the \cpp{IntegratedLuminosityNormalization} simulation item, and the edit fields below correspond to floating point properties. The user has selected the `custom wavelength range' option and has entered values in the edit fields appropriate for the problem under consideration. If the user would select the `all wavelengths' option instead, the two fields specifying the wavelength range would disappear because they are irrelevant for that option. The text in the box at the lower left indicates the context of the first property currently shown in the window. In this case, the luminosity normalization is being configured for a \cpp{GeometricSource} held by the \cpp{SourceSystem}, which in turn is held by the top-level \cpp{MonteCarloSimulation}. This information is particularly relevant for advanced users who wish to locate the property in the corresponding XML configuration file or even in the source code.

Apart from a more appealing visual approach, the MakeUp wizard offers several benefits over the command-line Q\&A session. Key features include the ability to move backwards to the preceding question(s) and the option to open and modify an existing configuration file. There also is context-sensitive help on the items and properties being configured through built-in hyperlinks to the SKIRT web site.

\subsection{Providing reliability statistics}
\label{Statistics}

\begin{figure*}[t]
  \centering
  \includegraphics[height=0.58\textwidth]{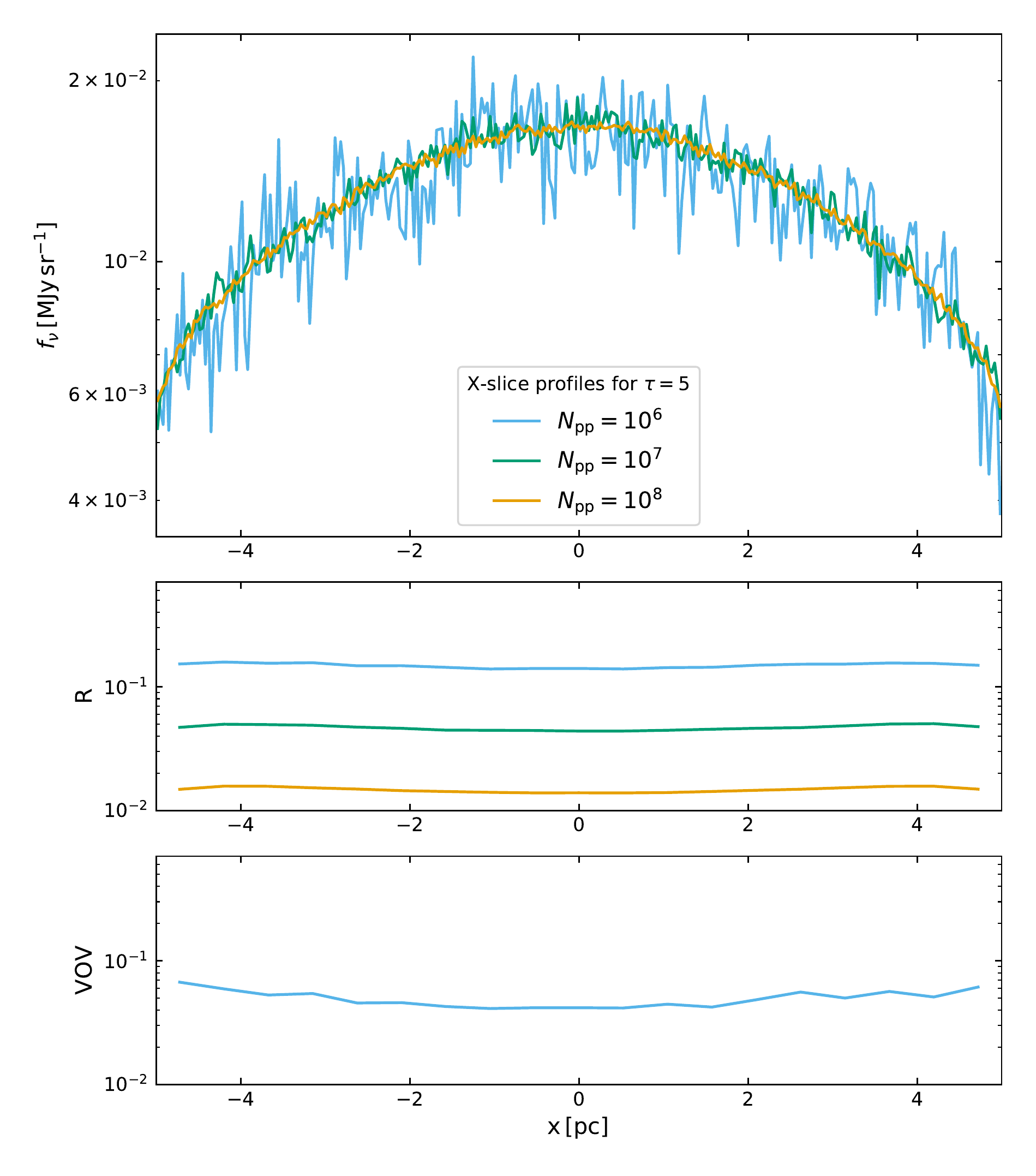}%
  \includegraphics[height=0.58\textwidth]{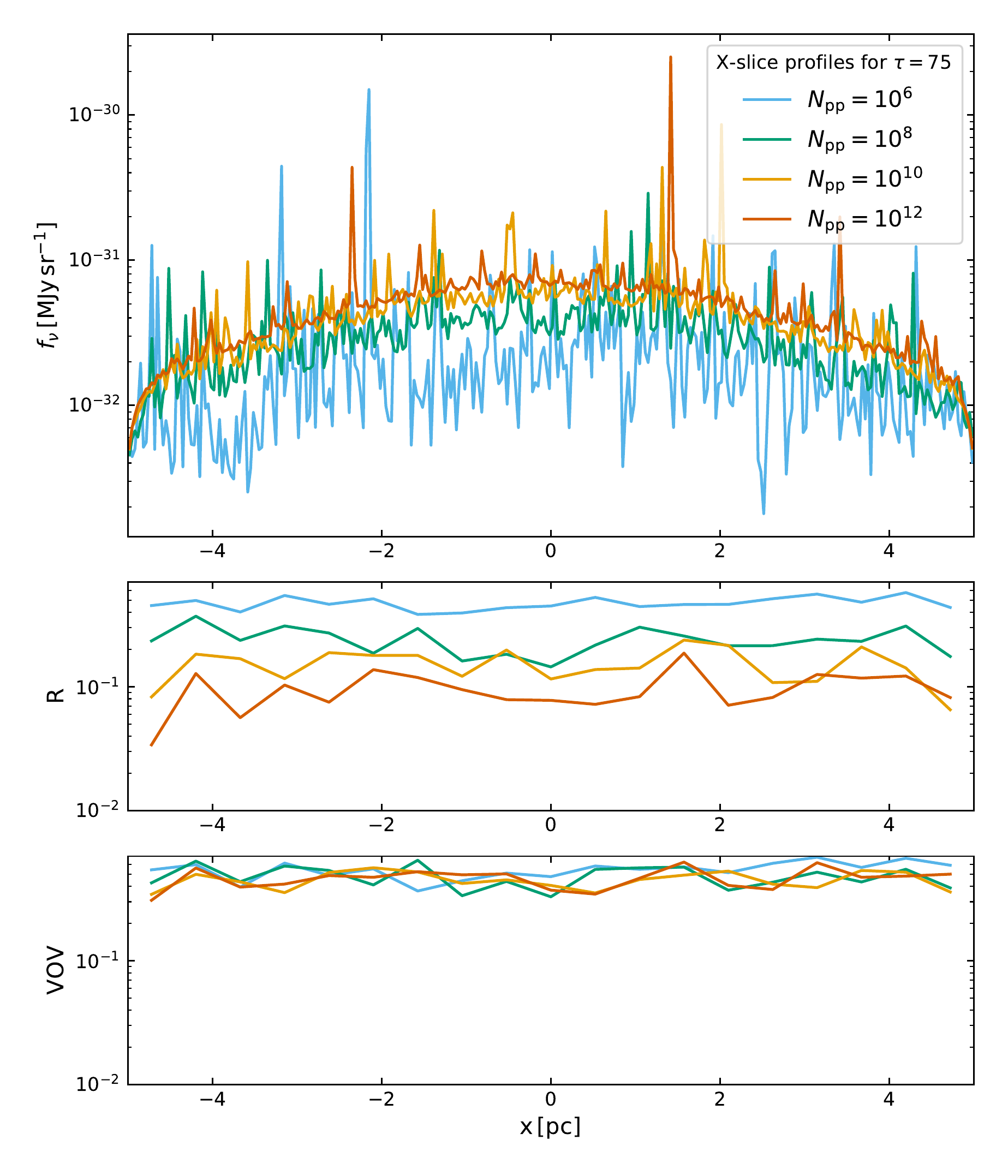}
  \caption{The flux density detected by SKIRT along a slice of the 3D slab defined by \citet{Gordon2017} on the side opposite from the source (top row) and the corresponding values of the relative error $R$ (middle row) and the variance of the variance VOV (bottom row). The data points for the latter two rows were binned to reduce scatter. The left panel shows results for an optical depth across the slab of $\tau=5$, the right panel for $\tau=75$. The different colors show results for a varying total number of photon packets launched during the simulation. The VOV curves in the lower left panel for $N_\mathrm{pp}\geq10^7$ are missing because all values are below $10^{-2}$. }
  \label{fig:slabmono}
\end{figure*}

Because of the probabilistic nature of the technique, the output quantities of a Monte Carlo radiation transfer (MCRT) simulation code such as SKIRT are inherently subject to noise. Generally, the signal to noise ratio can be improved by tracing a larger number of photon packets. Basic statistical treatments have been proposed to obtain a confidence interval in the context of astrophysical dust MCRT simulations by, e.g., \citet{Gordon2001} and \citet{Steinacker2013}. However, these techniques seem to fall short for certain model configurations with high optical depth. \citet{Gordon2017} study a 3D benchmark model involving a point source and a cuboid-shaped dust slab. Comparing the results generated by several simulation codes for the radiation penetrating the slab for higher optical depths, they find a lack of agreement between the codes and a lack of convergence with increasing number of photon packets even for the same code. Specifically, using SKIRT 8 on this benchmark model, \citet{Camps2018b} report that for an optical depth of $\tau\approx 75$ across the slab both the detected intensity profile along the slab and the integrated flux change substantially with the number of photon packets launched into the slab, even if the noise level in the intensity profile seems acceptable.

\citet{Camps2018b} define a simplified, 1D plane-parallel version of the slab configuration that allows accurate reference solutions to be calculated using a deterministic method. They reproduce the lack of convergence of the MCRT method for radiation penetrating the plane-parallel slab at optical depths similar to those reported for the 3D slab. Following earlier work in the field of nuclear particle transport simulations \citep[see, e.g.,][]{Pederson1997, MCNP2003}, they describe more advance statistical tests and show their effectiveness for the 1D slab problem. In their conclusion, \citet{Camps2018b} argue that MCRT codes should be equipped with these statistical mechanisms so that users can evaluate the accuracy of the simulation results. This is what we decided to implement in SKIRT~9.

We first summarize the relevant definitions, following \citet{Camps2018b} and the earlier work mentioned in the previous paragraph. The total SKIRT simulation result for a given flux recorder or instrument bin (see Sect.~\ref{Output}) is obtained by accumulating the contributions $w_i$ of individual photon packets, i.e.\ $\sum_i w_i$, where the index $i$ and the sum run over the $N$ photon packets launched during the simulation. To achieve proper statistics, all contributions to the same bin from the complete scattering history of a given photon packet launch are in fact aggregated into a single contribution $w_i$. If so requested, SKIRT will track the sums $W_k=\sum_i w_i^k$ for $k=0,1,2,3,4$ during the simulation (as opposed to just the sum for $k=1$) and output these data alongside the regular results. The sum $W_0$ yields the number of photon packet contributions to the given bin, which by itself may be an interesting statistic. Using the higher order sums one can estimate the relative error, $R$, as
\begin{linenomath}\begin{equation}
R = \left[ \frac{W_2}{W_1^2} - \frac{1}{N} \right]^{\nicefrac{1}{2}}
\end{equation}\end{linenomath}
and the variance of the variance, VOV, as
\begin{linenomath}\begin{multline}
\textstyle \mathrm{VOV} = \left[ W_4 - 4 W_1 W_3/N + 8 W_2 W_1^2/N^2 \right. \\
\textstyle \left. - 4 W_1^4/N^3 - W_2^2/N \right] \left. \middle/ \right. 
\left[ W_2 - W_1^2/N \right]^2
\end{multline}\end{linenomath}
The authors of the earlier work mentioned above recommend that the relative error $R$ should be smaller than $0.1$ for the corresponding result to be considered reliable. In that case, $R$ can indeed be interpreted as a relative error on the result. In the range $0.1<R<0.2$, however, results are questionable, and for $R>0.2$, results are unreliable. The VOV is a quantity that, in turn, measures the statistical uncertainty in the value for $R$. It is much more sensitive to large fluctuations in the $w_i$ values than is $R$, and it can thus detect situations where the obtained $R$ value is unreliable. The value of the VOV should be below 0.1 to ensure a reliable confidence interval.

To see these statistics at work in a 3D context, we configure a SKIRT 9 simulation with the 3D slab geometry defined by \citet{Gordon2017} using a single, optical wavelength and a single instrument on the side of the slab opposite to the source ($\theta=180^\circ$). The instrument frame covers a field of view across the length of the slab in the X direction and a slice of about 10\% of the slab width in the Y direction; it is offset from the center so that it matches the X slice defined in figure 7 of \citet{Gordon2017}. The number of pixels is set to $300\times1$ so that each pixel spans the full width of the instrument frame, and all photon packet contributions along the width of the slice end up in the same bin. Fig.~\ref{fig:slabmono} shows the flux density profile detected by this instrument for a varying total number of photon packets $N_\mathrm{pp}$ launched during the simulation (top row) and the corresponding values of the relative error $R$ (middle row) and the variance of the variance VOV (bottom row).

The left panel of Fig.~\ref{fig:slabmono} implies that for an optical depth across the slab of $\tau=5$ the results are essentially converged starting at $N_\mathrm{pp}\geq10^7$. In this case, $R<0.1$ and the value of the VOV is below $10^{-2}$ (which is why the corresponding curves are missing), so that both $R$ and the actual simulation results can be deemed reliable. The situation shown in the right panel of Fig.~\ref{fig:slabmono} for an optical depth of $\tau=75$ is totally different. The average flux density rises significantly with an increasing number of photon packets, in line with the results reported by \citet{Gordon2017} and \citet{Camps2018b}. The good news is that we can now use statistical analysis to evaluate the results. It appears that $R$ (finally) dips under the reliability threshold for $N_\mathrm{pp}=10^{12}$. However, the VOV is still well above the threshold, so the value of $R$ cannot really be trusted. This is reflected by two features apparent in the flux density profiles. Firstly, extrapolation of the progression of the flux profile average with increasing $N_\mathrm{pp}$ seems to indicate room for a further increase. Secondly, the flux density curve shows several large peaks of an order of magnitude or more even for the simulation with $N_\mathrm{pp}=10^{12}$. We can conclude from this analysis that the results are still not fully converged. 

The simulation with $N_\mathrm{pp}=10^{12}$ shown in the right panel of Fig.~\ref{fig:slabmono} ran for over 6 hours on a 24-core server. Calculating the next step in the progression, i.e.\ $N_\mathrm{pp}=10^{14}$, would take 25 days on the same machine. While using a larger, multi-node computer system could shorten the elapsed time, there is no guarantee of convergence. In fact, the lack of progression of the VOV values up to $N_\mathrm{pp}=10^{12}$ seems to indicate that we would need to process several orders of magnitude more photon packets for the results to be converged with an acceptable signal to noise ratio.


\section{Validation and performance}
\label{Performance}

\begin{figure*}
\centering
\begin{minipage}[c]{0.93\textwidth}
\begin{lstlisting}[style=cpp]
Instruments/Angles: Succeeded
Instruments/Calibration: Succeeded
Instruments/ComponentsA: Failed
  Differing files:
    plum_i1_sed.dat                       --      3 ( 50.00%) >0        2 ( 33.33%) >10        0 (  0.00%) >50
    plum_i1_total.fits                    --    892 (100.00%) >0      865 ( 96.97%) >10      790 ( 88.57%) >50
    plum_i2_primarydirect.fits            --      3 (100.00%) >0        0 (  0.00%) >10        0 (  0.00%) >50
    plum_i2_primaryscattered.fits         --    892 (100.00%) >0      867 ( 97.20%) >10      790 ( 88.57%) >50
    plum_i2_sed.dat                       --     12 ( 80.00%) >0        5 ( 33.33%) >10        0 (  0.00%) >50
    plum_i2_total.fits                    --    892 (100.00%) >0      865 ( 96.97%) >10      790 ( 88.57%) >50
    plum_i2_transparent.fits              --      3 (100.00%) >0        0 (  0.00%) >10        0 (  0.00%) >50
Instruments/ComponentsB: Failed
  Differing files:
    plum_i_primarydirect.fits             --     76 (100.00%) >0        5 (  6.58%) >10        0 (  0.00%) >50
    plum_i_primaryscattered.fits          --   5827 ( 99.73%) >0     5429 ( 92.91%) >10     4464 ( 76.40%) >50
    plum_i_secondarydirect.fits           --   4697 (100.00%) >0     4359 ( 92.80%) >10     3256 ( 69.32%) >50
    plum_i_secondaryscattered.fits        --    104 (100.00%) >0      104 (100.00%) >10      104 (100.00%) >50
    plum_i_sed.dat                        --    409 ( 80.35%) >0      181 ( 35.56%) >10      106 ( 20.83%) >50
    plum_i_total.fits                     --  10478 ( 99.85%) >0     9716 ( 92.59%) >10     7665 ( 73.04%) >50
    plum_i_transparent.fits               --     76 (100.00%) >0.       5 (  6.58%) >10        0 (  0.00%) >50
Instruments/FieldOfView: Succeeded
Instruments/WavelengthGrid: Succeeded
\end{lstlisting}
\end{minipage}
\caption{Excerpt from a report produced by the PTS~9 procedure performing functional tests. For failed tests, the report includes a list of output files that differ from the corresponding reference file with some extra statistics. The three columns list the number and percentage of nonzero values in the file that differ from the reference by more than 0, 10 and 50 percent, respectively.}
\label{fig:testreport}
\end{figure*}

\subsection{Functional tests}
\label{FunctionalTests}

Testing a complex code such as SKIRT is a nontrivial undertaking that requires constant care. There is the obvious need to evaluate new features as they are developed. It is also crucial to avoid breaking the existing functionality by code updates, no matter how large or small. And finally, simulation results must be validated in some objective and quantitative way. We have established two separate test procedures to address these needs: functional tests, introduced in this section, and benchmark tests, discussed in Sect.~\ref{Benchmarks}.

The \emph{unit} test concept is well known in the software development community. The \emph{functional} tests developed for SKIRT resemble unit tests, although there are significant differences. Unit tests operate from within the code, so that they can, in principle, access and test each and every code path. The functional tests for SKIRT operate from the outside, using SKIRT's flexible configuration capability to access many, but not all code paths. Specifically, they cannot test features such as optional command line arguments or the Q\&A user interface. For reasons explained later in this section, all functional tests are run in serial mode, so they cannot test SKIRT's parallelization features. Finally, each functional test by necessity relies on nontrivial sections of common code, such as the module loading the configuration file, while a unit test usually focuses on a specific small area in the code.

The Python toolkit for SKIRT (PTS, see Sect.~\ref{PythonToolkit}) includes a facility for running and validating a batch of functional SKIRT test cases. The test case definitions reside in a nested directory hierarchy; each test case consists of a particular SKIRT configuration file, optional input files, and a set of reference output files. The PTS procedure automatically locates and runs the tests, compares the generated output files to the reference files, and produces a concise report with a success/failure status for each test (see Fig.~\ref{fig:testreport}). For failed tests, the report also provides some details on the differences between the generated and reference output.

A major issue in this context is that the values in most SKIRT output files depend on the (pseudo-)random sequence used by the Monte Carlo processes throughout the simulation. For a sufficiently large number of photon packets, the results should be statistically equivalent. This is, however, hard to verify in an automated fashion and, perhaps more importantly, the run time for test cases should be kept at a minimum. So we need to ensure that the reference and test simulations are performed with the same random sequence.

Each execution thread in SKIRT is equipped with its private random number generator. Each of these generators receives a different seed at the start of the simulation. Subsequent SKIRT invocations will seed the generators in the same way, unless otherwise specified in the configuration file. However, parallel execution threads will receive chunks of work in an unpredictable order (see Sect.~\ref{Parallel}), causing the random number sequences used for each chunk to be mixed up unpredictably. Conversely, in serial mode, i.e.\ using a single process with a single execution thread, the random number sequence will always be the same. Therefore, the PTS procedure always performs each test in serial mode, launching multiple SKIRT instances for different tests in parallel to optimally use the available computing resources.

One important use of the automated tests is to verify the operation of the code after an update. If a code adjustment causes, for example, an additional random number to be consumed at some point in time, all subsequent calculations and results will change (although they should be statistically equivalent). In such situations, a human must evaluate the new test results and update the reference output where appropriate. To assist the developer with this verification, the automated test procedure provides some statistics on the changes to each file. As shown in Fig.~\ref{fig:testreport}, the report includes three columns listing the number and percentage of nonzero values that differ by more than 0, 10 and 50 percent, respectively.

Another problem is that the precise output of a numeric calculation may vary between run-time environments. As a first example, the evaluation order of function arguments in C++ is unspecified (C++ Standard, section 5.2.2/8) and thus sometimes differs between compilers. When two or more of the arguments to the same function call request a random number, the random number sequence used in the function body will differ between compilers. This is easily avoided by requesting the random numbers in separate statements ahead of the function call. As a second example, the result returned by the cosine and sine functions sometimes differs in the least significant bit between implementations of the standard library. There seems to be no solution to this problem other than creating and manually verifying a set of reference files for each relevant operating system/compiler combination.

Despite these limitations, we believe to have created a powerful suite of functional tests that cover most of the relevant features and code paths. Whenever a new SKIRT feature is developed, one or more test cases are added, and the reference output files are verified by the developer. As a side benefit, the requirement of developing these tests often acts as a reminder to also test marginal cases. At the time of writing, there are more than 400 test cases. Because the full test suite completes in just a few minutes on a present-day multi-core desktop computer, it is quite feasible to run it before a code update is committed to the master branch in the GitHub repository. So far, our functional tests have already captured, and thus prevented, quite a few regression bugs that would otherwise have gone unnoticed at least initially.

\subsection{Benchmarks}
\label{Benchmarks}

While the functional test suite discussed in the previous section is a formidable tool for verifying specific features and avoiding regression issues, there is also a need for validating simulation results for more realistic and complex models. Because nontrivial radiative transfer problems cannot be solved analytically or with deterministic numerical methods, the only option is comparing the results of different codes. This realization has led several authors to present well-defined benchmark problems with corresponding reference solutions produced by a number of codes participating in the benchmark.

SKIRT~9 successfully performs the relevant dust radiation transfer benchmarks available in the literature, as summarized below. The geometries, source spectra and dust properties needed for these benchmarks are built into the code. The `benchmark' section of the SKIRT web site shows the full results for each benchmark, and offers the corresponding configuration files for download, so that any interested third party can run the benchmark simulations. 

Unless otherwise noted below, the SKIRT results are within the uncertainty levels implied by the variations in the solutions generated by the codes participating in the benchmark.

\paragraph{Spherically symmetric circumstellar dust shell \citep{Ivezic1997}} This benchmark uses a 1D geometry consisting of a star embedded in a spherical dust shell with different dust density profiles and optical depths up to $\tau(1\mu\mathrm{m})=1000$. SKIRT does not properly handle the case with the highest optical depth for the steepest density gradient. This is related to the discussion in Sect.~\ref{Statistics} about radiation penetrating a slab with high transverse optical depth. The MCRT method simply does not seem to handle such problems well, while the codes participating in the original \citet{Ivezic1997} benchmark used 1D methods which could properly treat high optical depths. As noted by \citet{Gordon2017}, the 2D disks in the benchmarks described in the following two paragraphs do not suffer this problem because they have limited optical depths along the rotation axis of the model, so that the global scattered flux is dominated by scattering at low optical depths rather than at the high disk plane optical depths.

\paragraph{Axisymmetric circumstellar dust disk \citep{Pascucci2004}} This benchmark has a 2D geometry consisting of a star embedded in a circumstellar disk, for edge-on optical depths up to $\tau(0.55\mu\mathrm{m})=100$. SKIRT properly performs this complete benchmark.

\paragraph{Polarization from scattering in an axisymmetric dust disk \citep{Pinte2009}} This benchmark has a 2D geometry similar to that of the one presented in the previous paragraph, now including the effects of polarization for anisotropic scattering by spherical dust grains. SKIRT properly produces intensity and polarization maps at wavelength $\lambda=1\,\mu\mathrm{m}$ for a disk with edge-on optical depth $\tau(1\mu\mathrm{m})=45\times10^4$.

\paragraph{Stochastically heated dust grains \citep{Camps2015b}} This benchmark tests the calculation of the emission by stochastically heated dust grains, which plays an important role in the radiative transfer problem for a dusty medium. SKIRT participated in this benchmark effort.

\paragraph{Dust slab externally illuminated by a star \citep{Gordon2017}} This 3D benchmark includes a uniform dust slab externally illuminated by a star. The aim is to test dust absorption, scattering, and emission, optionally taking into account stochastic heating of dust grains. The authors provide results for transverse optical depths up to $\tau(1\mu\mathrm{m})=10$ and for several viewing angles. SKIRT participated in this benchmark effort. For a discussion of high slab optical depths, also see Sect.~\ref{Statistics} and Fig.~\ref{fig:slabmono}.

\paragraph{Polarization test cases \citep{Peest2017}} The authors define a number of basic test configurations to verify the effects of scattering by spherical dust grains on the polarization of radiation by comparison with analytically calculated solutions. SKIRT properly performs these test cases.

\subsection{Comparison to SKIRT~8}
\label{Comparison89}

\begin{figure}
  \centering
  \includegraphics[width=0.85\columnwidth]{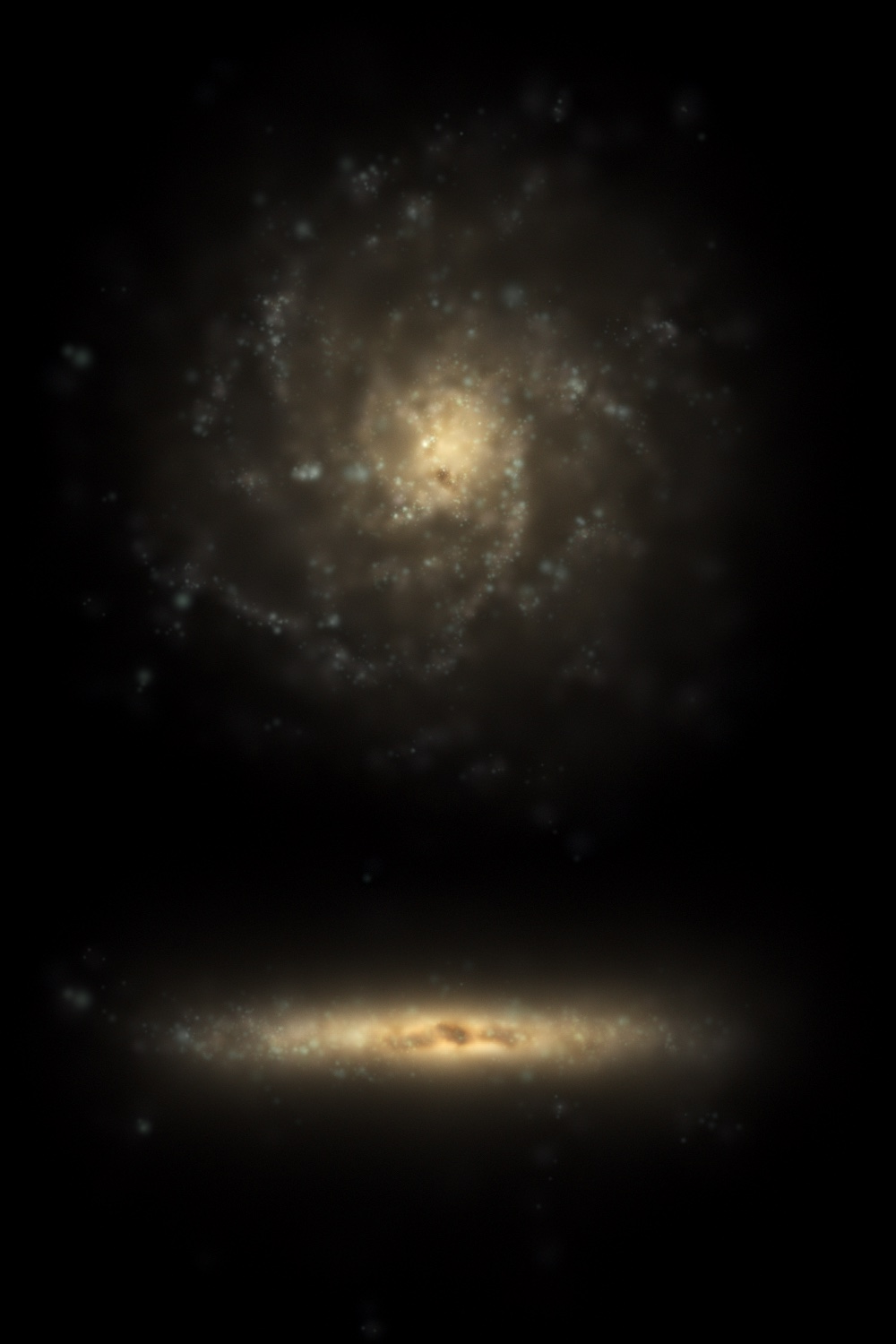}
  \caption{Face-on and edge-on views of a selected artificial galaxy \citep[galaxy ID 639646 from EAGLE simulation RefL0025N0752, ][]{Schaye2015,Crain2015} post-processed as described by \citet{Camps2016}. The illustration \citep[taken from the cover of][]{Camps2016phd} mimics an optical view as it would be seen by the human eye, combining the SDSS $g$, $r$ and $i$ bands. The simulated galaxy has a stellar mass of $1.75\times 10^{10}~\mathrm{M}_\odot$ represented by more than $125000$ stellar particles, and a dust mass of $3.8\times 10^{7}~\mathrm{M}_\odot$ derived from over $20000$ gas particles.}
  \label{fig:galaxyoptical}
\end{figure}

\begin{figure}
  \centering
  \includegraphics[width=0.99\columnwidth]{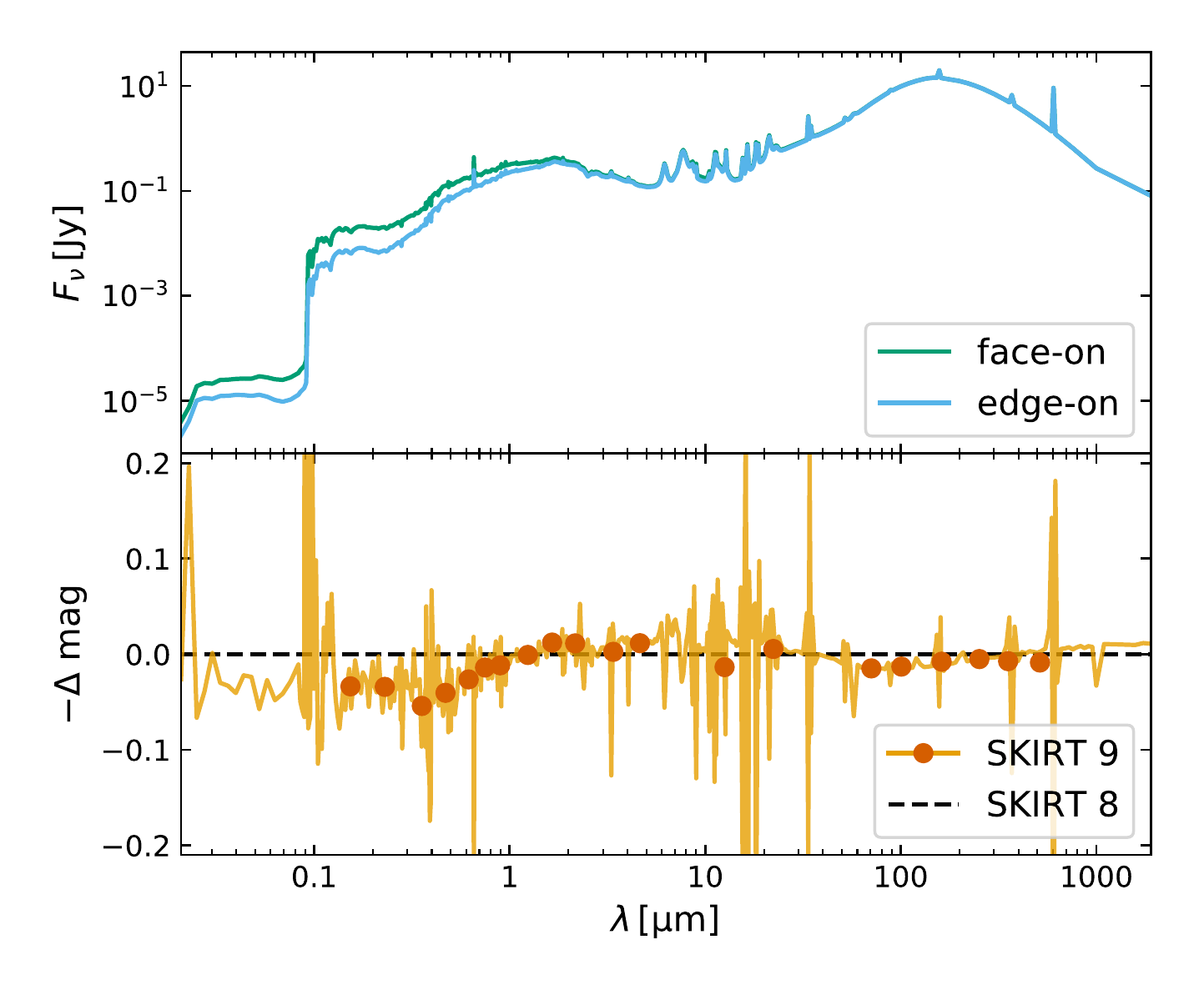}
  \caption{Top: emission spectra calculated by SKIRT~8 for the face-on and edge-on views of the artificial galaxy shown in Fig.~\ref{fig:galaxyoptical}. Bottom, solid line: difference between the edge-on spectra calculated by SKIRT~8 and SKIRT~9; dots: corresponding convolved flux density differences for some well-known observatories (GALEX, SDSS, 2MASS, WISE, and Herschel). }
  \label{fig:galaxysed89}
\end{figure}

We already compared many SKIRT~9 results directly or indirectly to those produced by SKIRT~8, for example while developing the functional test cases (see Sect.~\ref{FunctionalTests}) and performing the benchmark tests (see Sect.~\ref{Benchmarks}). In this section, we focus on comparing SKIRT~8 and SKIRT~9 results for a fairly sophisticated input model that is part of an actual science case, i.e.\ the well-resolved artificial disk galaxy illustrated in Fig.~\ref{fig:galaxyoptical} and described in the figure's caption. The number of particles representing this galaxy (over $125000$ stellar particles and over $20000$ gas particles) is sufficiently large for the radiative transfer simulation to be nontrivial, without being a limiting factor for running several tests with varying configuration parameters. To properly resolve the spatial structure of the input model, we configured a spatial octree grid \citep{Saftly2013} with approximately $1.14$ million cells in both SKIRT~8 and SKIRT~9.

We calculate the observed SED for the face-on and edge-on views of the artificial galaxy (see top panel of Fig.~\ref{fig:galaxysed89}). We focus our analysis on the edge-on SED because the more prominent dust extinction along this line of sight poses a more challenging radiation transfer problem. We record the flux densities on the 450-point wavelength grid presented by \citet{Camps2016}. In SKIRT~8 this is in fact the global wavelength grid in the simulation; in SKIRT~9 it is configured as the instrument wavelength grid. The other wavelength grids in SKIRT~9 (see Sect~\ref{Wavelengths}) and the number of photon packets launched in both SKIRT~8 and SKIRT~9 are configured to obtain converged results. This is defined to mean that increasing the number of grid points or number of photon packets changes the calculated flux densities by less than one per cent or, equivalently, 0.01 mag.

Next to the SED tabulated on a 450-point wavelength grid, we also calculate flux densities for the broadbands defined by some well-known observatories (GALEX, SDSS, 2MASS, WISE, and Herschel). For SKIRT~8, the convolution with each transmission curve is calculated after the fact from the tabulated SED. For SKIRT~9, the convolution occurs on the fly during the radiative transfer simulation (see Sect.~\ref{Bands}) and the results are placed in a separate table.

The bottom panel of Fig.~\ref{fig:galaxysed89} compares the SKIRT~8 and SKIRT~9 results. For most of the wavelength range, the correspondence is better than $\pm0.1$~mag, equivalent to $\pm10\%$. However, in areas where the SED varies rapidly, the differences can get more extreme. For example, the results near the 912~nm Lyman limit differ by nearly 3~mag, or more than a factor of 10 (well outside the range of the plot). Convolution with the broadband transmission curves smooths the variations, but a discrepancy of up to $\pm0.05$~mag remains.

The reasons for these differences are related to the unavoidable discretization of the physical processes. To begin with, distributing the source luminosity over a finite number of photon packets introduces noise. For our simulations, however, the convergence tests described earlier in this section ensure that this noise is limited to $\pm0.01$~mag. The true reason is the different handling of wavelengths in both codes (see Sect.~\ref{Wavelengths}). For each bin in the instrument wavelength grid, SKIRT~8 samples the source spectrum at just a single point (the center of the bin), while SKIRT~9 samples at many arbitrary wavelengths, essentially performing a Monte Carlo integration over the bin at the full resolution of the source spectrum. If the source spectrum (or the extinction by the medium) show strong variation within the wavelength bin, this can make a substantial difference. Furthermore, depending on the precise form of the spectrum, the difference will vary with the width of the instrument wavelength bin. 

\begin{figure}
  \centering
  \includegraphics[width=0.99\columnwidth]{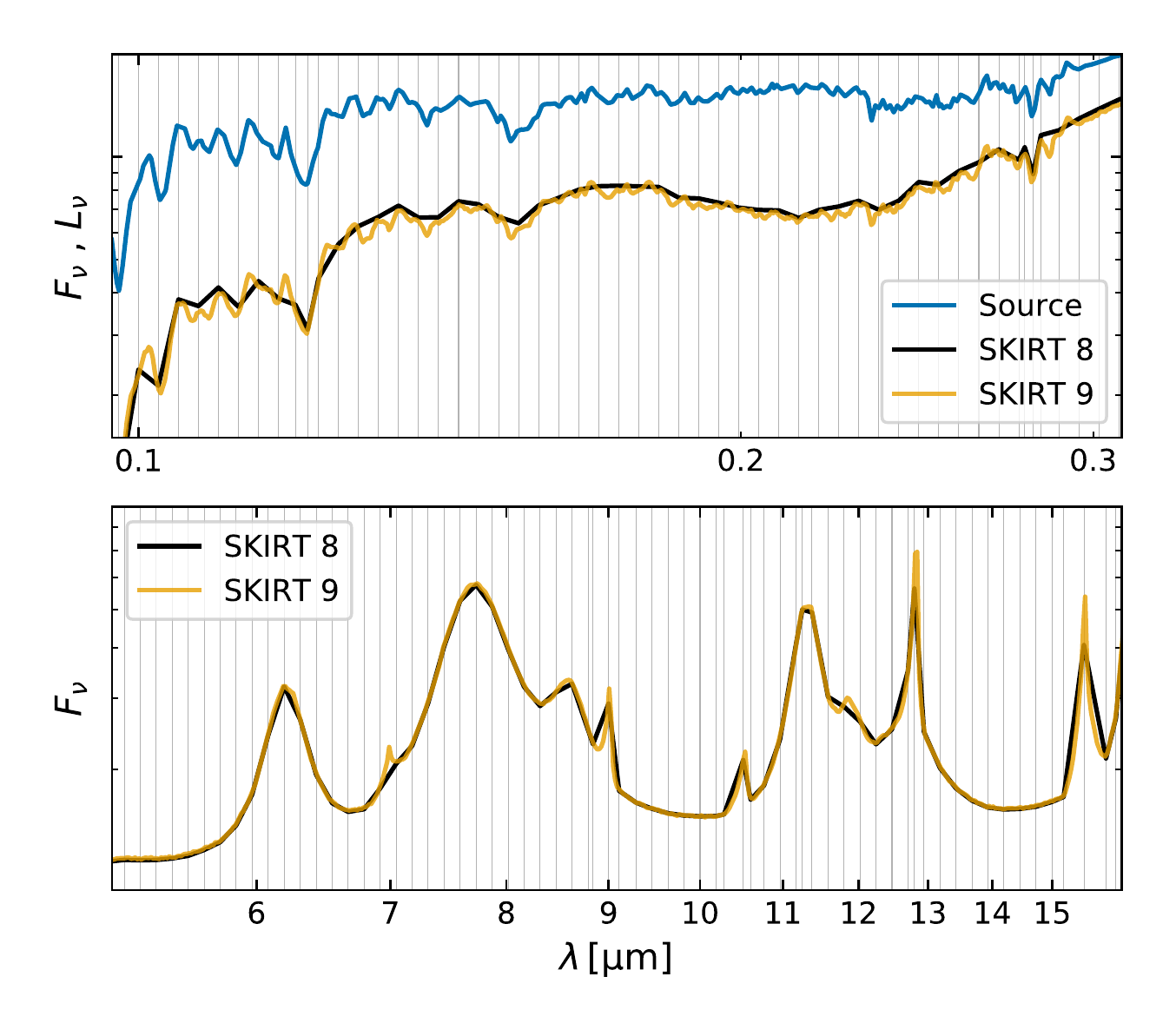}
  \caption{Segments of the emission spectrum for the edge-on view of the artificial galaxy shown in Fig.~\ref{fig:galaxyoptical}. The SKIRT~8 configuration is identical to that used for Fig.~\ref{fig:galaxysed89}. The 450-point wavelength grid is indicated by the vertical lines in the figure, with bin widths of mostly 0.01 dex in the ultraviolet range (top panel) and 0.008 dex in the infrared range (bottom panel). The SKIRT~9 configuration is also the same, except that it uses a high-resolution instrument wavelength grid with bin widths of 0.001 dex everywhere. The top panel also shows the stellar source spectrum at its full native resolution for comparison.}
  \label{fig:galaxyhires89}
\end{figure}

We argue that the SKIRT~9 results are more accurate. The SED templates employed in our simulation for representing the stellar populations have a higher spectral resolution than the instrument wavelength grid, especially in the ultraviolet/optical wavelength range. SKIRT~9 integrates over the spectrum at its native resolution, while SKIRT~8 does not. To illustrate this, the top panel of Fig.~\ref{fig:galaxyhires89} shows a zoom on the ultraviolet portion of Fig.~\ref{fig:galaxysed89}, using arbitrary, logarithmic units on the vertical axis. The topmost solid line traces the spatially integrated emission spectrum of the stellar sources in the galaxy, ignoring extinction, and is plotted with full native spectral resolution. The SKIRT~8 SED is actually a straightforward zoom-in of the edge-on spectrum shown in Fig.~\ref{fig:galaxysed89}. It has an extra slope because of the wavelength-dependent dust extinction. The vertical lines indicate the 450-point wavelength grid used by SKIRT~8. It is apparent that the form of the source SED is sampled on the wavelength grid points. The SKIRT~9 SED is produced using a high-resolution instrument wavelength grid (about 500 points across the wavelength range shown in the figure). Except for replacing the instrument wavelength grid, the simulation configuration is identical to the one used for Fig.~\ref{fig:galaxysed89}, including, for example, the number of photon packets. This SED traces the form of the source SED in much greater detail, and as described, SKIRT~9 preserves this information even if the bins of the instrument wavelength grid are much wider. This effect most certainly contributes to the seemingly systematic discrepancies of Fig.~\ref{fig:galaxysed89} in the ultraviolet/optical wavelength range.

Similarly, our SKIRT~9 simulation has been configured to calculate dust emission spectra with higher resolution than what is feasible in SKIRT~8, and it again automatically `integrates' over the full resolution of this spectrum. This is illustrated in the bottom panel of Fig~\ref{fig:galaxyhires89}, where it is again apparent that the SKIRT~9 SED shows more detail in the emission features.

When considering broadband fluxes, the on-the-fly convolution mechanism of SKIRT~9 additionally bypasses the discretization step occurring at the instrument side in SKIRT~8, namely binning the recorded fluxes into the instrument wavelength grid before the convolution can be performed. This is especially relevant for instruments that record image frames, because the per-pixel noise levels are usually a lot higher than those for spatially integrated fluxes.

\subsection{Wavelength grid configuration and convergence}
\label{WaveConfig}

As discussed in Sects.~\ref{Wavelengths} and \ref{Comparison89} and illustrated in Fig.~\ref{fig:wavelengths89}, SKIRT~9 uncouples the wavelength grids for instruments from the treatment of wavelengths in other areas of the code. One useful application of this flexibility is to configure two instruments for the same line of sight: an `SED instrument' for recording a high-resolution spectrum, and a `Frame instrument' for recording a small set of broadband images. Each instrument performs different binning on the same set of arriving photon packets: the SED instrument spatially integrates fluxes into narrow spectral bins, while the frame instrument does a spectral convolution for the fluxes in every spatial pixel. This leads to a very efficient use of the photon packets being traced through the system. To achieving a similar result with SKIRT~8, one would need to configure a single instrument that records a data cube with both the required spatial and spectral resolution, and perform the two-way binning after the fact. Apart from requiring an extra processing step, the three-dimensional data structures in this approach can become prohibitively large (see Sect~\ref{Memory}).

For a panchromatic simulation as the one described in Sect.~\ref{Comparison89}, SKIRT~9 requires the user to configure two wavelength grids that affect its internal operation. The first one is the `dust emission wavelength grid', which controls the resolution of the dust emission spectrum calculated for each spatial cell. Especially when taking into account the stochastic heating of small dust grains, this spectrum contains many narrow infrared features as shown in Figs.~\ref{fig:galaxysed89} and \ref{fig:galaxyhires89} \citep[and see also][]{Camps2015b}. It is thus desirable to configure a grid that can properly resolve these features. Memory is not an issue because the emission spectrum is stored just once per execution thread (see Sect.~\ref{DistributingPackets}). The performance impact is very limited as well because sampling these emission spectra is not the bottleneck of the calculation. For the simulations in Sect.~\ref{Comparison89}, we configured an emission wavelength grid with a resolution of 100 bins per dex in the overall range from $0.2$ to $2000~\mu\mathrm{m}$, and 200 narrower bins in the range from $3$ to $25~\mu\mathrm{m}$, for a total of 508 bins.

The second important internal wavelength grid is the `radiation field wavelength grid', which defines the bins used to record the energy deposited by photon packets in each spatial cell (see Sect.~\ref{RecordingField}). Generally, we expect the radiation field at shorter wavelengths to dominate the dust heating process, with the longer wavelengths having a minimal effect. We can also presume that the precise wavelength of an incoming photon packet might not be so important, as long as its energy is properly categorized, possibly allowing fairly wide wavelength bins. To investigate these issues, we performed convergence tests for the galaxy model discussed in Sect.~\ref{Comparison89}. As a reference we used a radiation field wavelength grid with 90 points from $0.02$ to $20~\mu\mathrm{m}$ and 20 more points from $20$ to $2000~\mu\mathrm{m}$. We finally settled on a grid with just 40 points from $0.02$ to $10~\mu\mathrm{m}$, distributed evenly in log space. The deviations in the spatially integrated SEDs (between simulations with these two grids) are smaller than one per cent, which is of the order of our convergence criterion for other discretization parameters. Likewise, the deviations in the calculated indicative dust temperatures for the spatial cells in the simulation are of the same order as the Monte Carlo deviations between two simulations that use the exact same configuration. We conclude that the 40-point radiation field wavelength grid is appropriate for this particular type of input model. This is an important result, because the memory requirements for a SKIRT simulation critically depend on this number of bins, as discussed in Sect.~\ref{Memory}. Note that, before adopting such a grid for other input models, convergence tests similar to the one described here should be performed.

\subsection{Memory usage}
\label{Memory}

We have alluded in Sects.~\ref{Parallel}, \ref{Bands}, and \ref{WaveConfig} that SKIRT~9 can be more memory efficient than SKIRT~8. In this section, we attempt to quantify this claim. There are obviously countless areas in the SKIRT code that consume memory. In many cases, however, the overall memory requirements are dominated by just a few components, namely the radiation field storage and the instrument data cubes.

Consider a typical panchromatic dust continuum simulation in SKIRT~8. We assume that the global wavelength grid has $N_\lambda$ points and that the spatial grid has $N_\mathrm{cell}$ points. The data structure for storing the radiation field has a size in bytes of
\begin{linenomath}\begin{equation}
S_\mathrm{rf} = 8 N_\lambda N_\mathrm{cell}.
\label{eq:datasize-rf}
\end{equation}\end{linenomath}
If the simulation supports dust self-absorption, it stores the absorbed energy from stellar and dust emission separately, so that the data structure has twice this size.

We further assume an instrument with $N_\mathrm{x}$ by $N_\mathrm{y}$ image pixels. The data structure for recording the detected fluxes has a size in bytes of
\begin{linenomath}\begin{equation}
S_\mathrm{ins} = 8 N_\lambda N_\mathrm{x} N_\mathrm{y}.
\label{eq:datasize-ins}
\end{equation}\end{linenomath}
If the instrument is requested to keep track of individual flux components (such as direct and scattered light), or if there are similar instruments at other viewing angles, the allocated data structures become a multiple of this size.

For a simulation with 500 wavelengths, 3 million spatial cells, and three 750 by 750 pixel instruments, the aggregate size of these data structures is 17.5~GB. For a more sizable simulation with 10 million cells and 1250 by 1250 pixel instruments, the number becomes 55~GB. This will grow even more when tracking individual flux components or including more instrument viewing angles. Specifically note that the total size scales proportionally with the number of wavelengths, because it is a multiplier in both Eqs.~(\ref{eq:datasize-rf}) and (\ref{eq:datasize-ins}). For example, performing a simulation with 5000 wavelengths would require 175~GB for the first example above, and 550~GB for the second example. This becomes prohibitive for all but the largest shared-memory computing systems, and it is why we introduced a data parallelization mode in SKIRT~8; see Sect.~\ref{Parallel} and \citet{Verstocken2017}.

Now consider the same type of panchromatic dust continuum simulation in SKIRT~9. Equations ~(\ref{eq:datasize-rf}) and (\ref{eq:datasize-ins}) remain valid, but the number of wavelengths $N_\lambda$ in each of the equations can now be different and is fully uncoupled from the spectral resolution required for spatially integrated spectra. Indeed, a pure SED instrument does not consume a significant amount of memory, even for a very large number of wavelengths. Assume that we configure a radiation field wavelength grid with 40 points as in Sect.~\ref{WaveConfig}, so that $N_{\lambda,\mathrm{rf}}=40$ in Eq.~(\ref{eq:datasize-rf}), and that we require image frames for 30 broadbands, so that $N_{\lambda,\mathrm{ins}}=30$ in Eq.~(\ref{eq:datasize-ins}). The total size of the data structures now becomes 1.3~GB for the first example above, and 4~GB for the second example, i.e.\ more than an order of magnitude smaller.

To be fair, it should be noted that some studies will need a spatially and spectrally resolved instrument data cube in a given wavelength range, for example to evaluate the effects of kinematics, or to simulate integral-field spectroscopy observations such as those made by MUSE \citep{Bacon2004} or SAMI \citep{Croom2012}. This affects $N_{\lambda,\mathrm{ins}}$ without changing $N_{\lambda,\mathrm{rf}}$ in the equations above. When we introduce other media types such as hydrogen gas, the wavelength resolution $N_{\lambda,\mathrm{rf}}$ of the radiation field storage will need to increase as well. In other words, at some future time we may need to revisit the possibilities for distributing data structures over multiple parallel processes in SKIRT~9.

\subsection{Processing time}
\label{Processing}

\begin{figure}
  \centering
  \includegraphics[width=0.99\columnwidth]{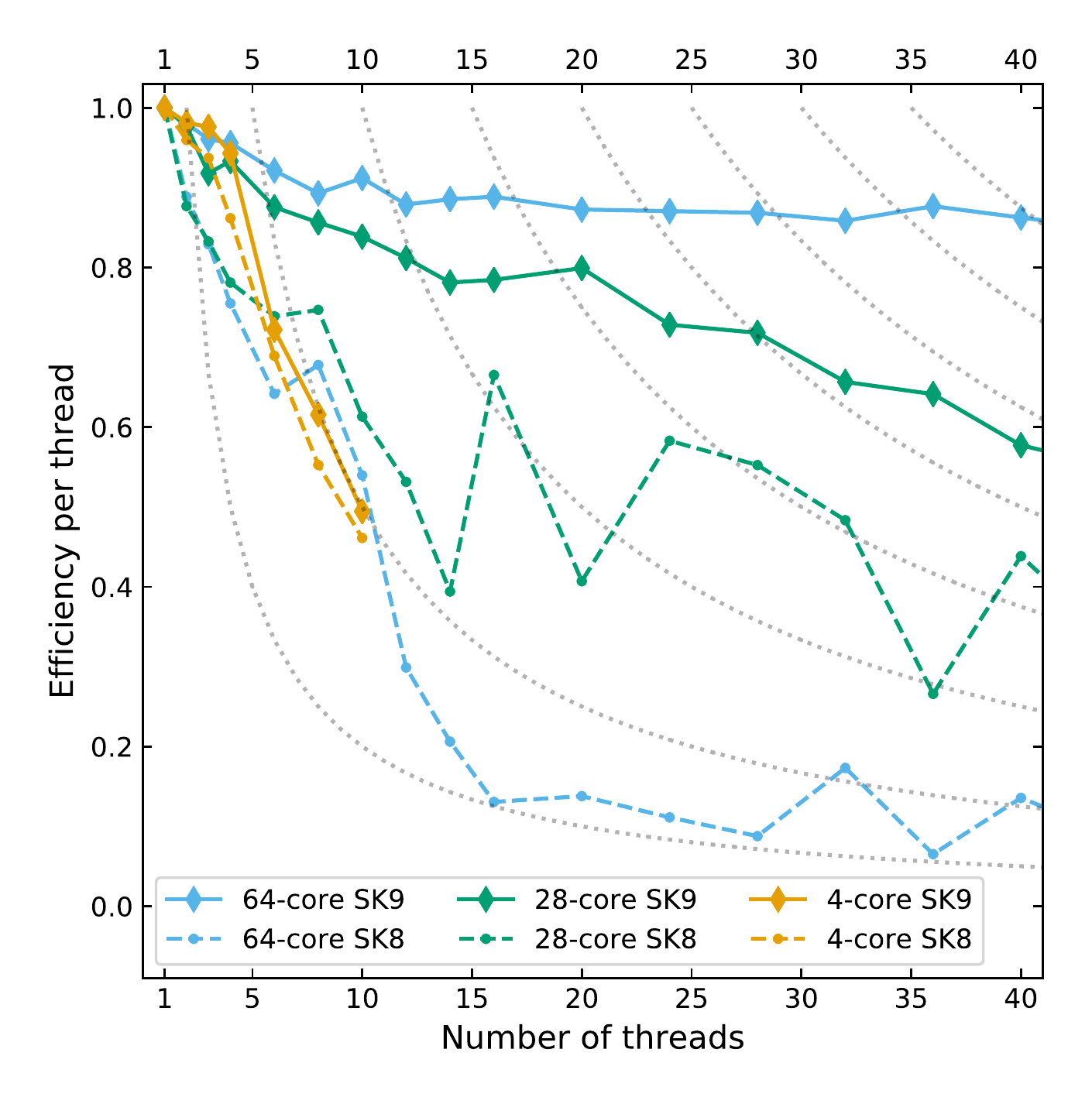}
  \caption{Scaling efficiency of parallel execution threads as defined by Eq.~(\ref{eq:scalingeff}) for the SKIRT~8 and SKIRT~9 primary emission phase. These results were measured on a recent desktop with 4 cores, a server with 28 cores, and an older server with 64 cores. Each of these systems has an equal number of extra `virtual' cores that share the hardware with the `true' cores. The black dotted lines indicate theoretical curves for $T(N)=T(1)/K$, i.e.\ perfect scaling for $K$ threads (where the curve starts at the top) and no benefit at all of adding threads beyond that.}
  \label{fig:parascaling}
\end{figure}

In this section we investigate the performance impact of the changes in SKIRT~9 relative to SKIRT~8. We begin by considering the scaling efficiency of multiple parallel execution threads in a single process for the primary photon packet emission loop. Given the number of parallel execution threads $N$ and the corresponding elapsed wall time $T(N)$, we define the efficiency per thread $E(N)$ as
\begin{linenomath}\begin{equation}
E(N) = \frac{T(1)}{N\,T(N)},
\label{eq:scalingeff}
\end{equation}\end{linenomath}
which is equal to one for perfect scaling and goes down as the efficiency of the parallel execution decreases. Fig.~\ref{fig:parascaling} shows measurements for SKIRT~9 (solid lines) and SKIRT~8 (dashed lines) on three computer systems with a varying number of cores. The black dotted curves trace lines of constant scaling, i.e.\ where $T(N)=T(1)/K$, or perfect scaling for $K$ threads and no benefit at all of adding threads beyond that.

It is evident from Fig.~\ref{fig:parascaling} that SKIRT~9 scales much better than SKIRT~8, especially for a larger number of threads. This is a consequence of our new load-distribution mechanism (see Sect.~\ref{Parallel}) requiring less inter-thread communication (for every chunk of photon packets as opposed to for every packet). The improvement is most extreme for the 64-core system in our test, where SKIRT~8 scales very poorly. In fact, above 12 threads, its performance degrades to the equivalent of 2 to 5 perfectly scaling threads, regardless of the number of threads being added. SKIRT~9 scales very well on this system; its efficiency stays well above 80\% for up to more than 64 cores, outside the range of the plot. The difference is much smaller on the 4-core system in our test, although SKIRT~9 still scales better. The 28-core system sits somewhere in between, and shows some surprising (actually rather reproducible) jumps in the curves. It is clear that the scaling behavior of our code(s) strongly depends on the computer system's architecture. We can assume that shared-memory systems with a larger number of cores will scale more poorly, because it becomes exceedingly hard to efficiently implement the required synchronization hardware. Unfortunately, as it so happens, the 64-core system in our test is several years older than the 28-core system, which is older than the 4-core desktop, so that we cannot verify this conjecture for hardware of the same generation.

Apart from the scaling behavior, absolute processing times also heavily depend on the computer system being used, and they sometimes even vary between subsequent runs on the same computer. In the remainder of this section, we therefore describe qualitative trends rather than providing precise quantitative comparisons. Our tests show that monochromatic simulations (using a single wavelength in the optical range) proceed at essentially the same speed in SKIRT~9 as compared to SKIRT~8. Even if serial (single-threaded) execution might be slightly slower, this is easily compensated by the improved parallel scaling. This means that the extended capabilities of SKIRT~9 have not significantly slowed down the basic operation of the code.

The picture becomes more ambiguous when we consider multi-wavelength or panchromatic simulations, possibly including dust emission. During the primary emission phase, photon packets are emitted from the (stellar) sources and traced through the spatial grid, while information about the radiation field is being recorded for each spatial cell, and peel-off photon packets are detected at each instrument. During the secondary emission phase, the dust emission spectrum is calculated for each spatial cell, and photon packets are subsequently emitted from these cells. According to our tests, depending on the configuration, the SKIRT~9 primary emission phase can be substantially slower and its secondary phase can be quite a bit faster than the corresponding SKIRT~8 phases. In some cases the difference can be up to a factor of three, but usually, and especially for larger production runs, it is a lot closer to one. We will now examine what causes these performance differences and when they occur.

A first issue affecting the performance of primary emission is assigning wavelengths and luminosities to new photon packets. In SKIRT~8 the wavelengths are simply determined by the user-configured global wavelength grid, and the luminosities are interpolated from the source spectrum at the resolution of that same global wavelength grid. SKIRT~9 instead randomly samples a wavelength from the source's spectral energy distribution or from the bias wavelength distribution (see Sect.~\ref{SamplingWavelengths}). The sampling procedure requires construction of the normalized cumulative probability distribution for the source spectrum at its full inherent resolution. For built-in geometrical sources with just a single SED, this construction occurs just once at the start of the simulation and so its performance impact is negligible. For imported sources, however, each particle or cell (called a `sub-source') is assigned a specific SED interpolated from a family of templates. These templates are often tabulated at high wavelength resolution, making the interpolation and construction of the cumulative distribution a time-consuming process. To minimize the performance impact, these data are preserved between photon packet launches from the same sub-source (see Sect.~\ref{DistributingPackets}). In other words, when the number of photon packets is much larger than the number of particles or cells, as is usually the case in a production run, the time spent on this extra calculation per sub-source is small relative to the total simulation time (although the wavelength sampling itself still impacts each photon packet). For test runs with just a few photon packets per sub-source, however, the relative performance impact can be a lot more extreme.

A second performance issue is the need to determine the bin index in various wavelength grids corresponding to a photon packet's wavelength at several occasions during the packet's lifetime. In SKIRT~8 this operation is trivial because the photon packet carries the bin index in the global wavelength grid as a property. In SKIRT~9, the photon packet carries an actual wavelength value rather than an index and the simulation uses many different wavelength grids for various purposes (see Sects.~\ref{Wavelengths} and Sects.~\ref{WaveConfig}), including grids for tabulating optical properties, storing the radiation field, and detecting photon packets in instruments. Because these grids can be irregular, the `conversion' from an arbitrary wavelength value to the corresponding bin index requires a binary search.\footnote{For a grid with equidistant border points in logarithmic space, the bin index is a linear expression in the logarithm of the wavelength. According to our tests, however, computing the logarithm of the wavelength is slower than performing a binary search unless the grid has many thousands of points.} Because there can be many different wavelength grids and because a photon packet's wavelength can change during its lifetime as a result of kinematic effects, the result of this conversion operation cannot be efficiently cached. Generally, the performance impact of this new wavelength treatment increases with the logarithm of the number of points in the various wavelength grids.

The secondary emission phase is not affected by these performance issues to the same degree. The dust emission wavelength grid usually has a fairly limited resolution so that the construction of the cumulative distribution and the subsequent wavelength sampling have a limited impact. Also, unless dust self-absorption has been enabled, the radiation field is not being recorded during the secondary emission phase, avoiding the binary search operation described in the previous paragraph. In fact, the secondary emission phase often runs faster in SKIRT~9 than in SKIRT~8. The main reason for this acceleration seems to be that SKIRT~9 calculates the emission spectrum for each spatial cell (or `sub-source') just before emitting photon packets from the cell and discards the information when all photon packets for the cell have been launched, very similar to the procedure for primary sources (see Sect.~\ref{DistributingPackets}). Apart from consuming much less memory, this keeps the data and the operations close to each other, while SKIRT~8 pre-calculates and stores all secondary emission spectra in a big data structure.

SKIRT~9 detects during setup whether the configuration includes any moving media. If so, it tracks the appropriate kinematic effects during the photon cycle as described in Sect.~\ref{Kinematics}. Apart from the need for calculating Doppler shifts at each photon packet interaction with the medium, this also means that some optimizations are no longer possible. For example, because the perceived wavelength depends on the velocity vector of the medium in each spatial cell, one can no longer assume a constant extinction cross section along the path of a photon packet. As a result, the run time of a simulation increases by a factor of up to two. Moving sources have a much smaller impact on performance because the extra work is limited to calculating the Doppler shift when launching each photon packet.

When comparing SKIRT~8 and SKIRT~9 run times, the total number of photon packets to be launched in each simulation phase is a relevant configuration parameter. If SKIRT~8 specifies $N_\mathrm{pp,8}$ packets for each of the $N_{\lambda,8}$ wavelengths in the global grid, a straightforward translation towards SKIRT~9 is to configure a total of $N_\mathrm{pp,9} = N_{\lambda,8} N_\mathrm{pp,8}$ packets. However, for a simulation with a wide wavelength range, this is not fully equivalent. The SKIRT~8 packets are distributed equally over all $N_{\lambda,8}$ wavelength points, while the SKIRT~9 packets are sampled from the source spectrum in combination with the bias distribution (see Fig.~\ref{fig:wavebias} and Sect.~\ref{SamplingWavelengths}). As a result it is hard to make precise comparisons. Usually, though, SKIRT~8 `looses' more photon packets to wavelength ranges that have an insignificant or no contribution to the simulation results. In other words, SKIRT~9 will often reach converged results with a smaller configured number of photon packets. 


\section{Summary and outlook}
\label{Summary}

We restructured our MCRT code SKIRT to support long-term objectives such as including gas media types in addition to dust, tracing emission and absorption lines in addition to continuum radiation, and modeling polarization by spheroidal dust grains aligned with magnetic field lines. In this paper, we presented the key design choices and how they impact the capabilities and performance of the code.

The most fundamental change is in the treatment of wavelengths throughout the code. Photon packets now carry an arbitrary wavelength value as opposed to an index into a discrete list of values, and separate wavelength grids can be configured for various purposes, such as tracking the radiation field, calculating thermal emission by the dust medium, and recording synthetic observations in each instrument. Apart from allowing Doppler shifts caused by interactions with moving media, this change also improves the accuracy of the simulation results. All source spectra are sampled at their full, inherent resolution and instruments can be equipped to record fluxes in one or more broadbands by performing on-the-fly convolution with the appropriate transmission curves. Uncoupling the wavelength grids for different purposes enables memory savings of up to an order of magnitude for dust continuum radiative transfer simulations of large models such as those imported from hydro-dynamical snapshots. The impact on processing time is limited, especially for `production runs', and is compensated at least in part by the improved distribution of photon packets over the wavelength ranges that are significant to the simulation results.

The simulation input model is now defined in terms of more generic concepts, such as \emph{sources} and \emph{media} instead of stellar components and dust components. Beyond these updated naming conventions, the model structure supports new features such as attaching velocity fields to sources and media, and can easily be extended to assign a grain alignment field to a dust medium component, for example, in support of including polarization by non-spherical grains.

We also described some selected features in more detail. We discussed the wavelength sampling procedures, the mechanisms for recording the radiation field and for handling broadband convolution in the instruments, a new feature providing statistics on the reliability of the calculated results, and an optional graphical user interface for configuring a simulation. These sections can inform interested users on the precise operation of the code and they may, hopefully, inspire the design of other codes.

SKIRT~9 properly runs over 400 handcrafted functional tests covering most of the code paths and it successfully performs all relevant benchmarks available in the literature. The updated code has a broad range of potential applications in its core field of dust continuum radiative transfer, now including kinematic effects. It provides an excellent platform for including additional physical processes, extending its potential application domain even further. Newly added features will automatically benefit from the existing infrastructure for defining and importing input models, generating output, and handling the Monte Carlo photon packet life cycle.

At the same time, we realize that such expansion of the code into new areas of expertise will require a collaborative effort involving users and developers outside of the core SKIRT team. We therefore increased our effort to make the source code and all documentation publicly available in a context that encourages collaborative development. We actively invite contributions from other authors in any form, including new scientific applications, problem reports, suggestions for improvements, and actual source code.

An immediate opportunity, using currently implemented features, is to study the effect of dust kinematics in galaxy models. We would also like to build improved sub-grid models for star formation regions and for ionized gas in the context radiative transfer simulations of synthetic galaxies imported from cosmological simulations. Over the coming years, we also hope to see projects implementing new physics in SKIRT, such as the polarization effects of spheroidal dust grains, Lyman-alpha line transfer, more general hydrogen line transfer (or aspects thereof), and X-ray radiation transfer. Readers who might be interested in participating in any of these projects are invited to contact us.


\section*{Acknowledgments}

We thank all SKIRT users and contributors for their help in giving shape to the code and to this paper.


\bibliography{skirt9}

\end{document}